# Using numerical-experimental analysis to evaluate rPET mechanical behavior under compressive stresses and MEX additive manufacturing for new sustainable designs

Jorge Manuel Mercado–Colmenero[1], M.Dolores La Rubia[2], Elena Mata-García[1], Moises Rodriguez-Santiago[1], Cristina Martin–Doñate[1*]

[1]Department of Engineering Graphics Design and Projects. University of Jaen. Spain

[2] Department of Chemical, Environmental and Materials Engineering,

University of Jaen, Jaen, Spain*Correspondence: cdonate@ujaen.es

Campus Las Lagunillas, s/n. Building A3-210 23071 Jaen (Spain) Phone: +34 953212821, Fax: +34 953212334

## Abstract

Purpose:

Because of the anisotropy of the process and the variability in the quality of printed parts, finite element analysis is not directly applicable to recycled materials manufactured using fused filament fabrication. This study investigates the numerical-experimental mechanical behavior modeling of the recycled polymer rPET manufactured by MEX (Material Extrusion) process under compressive stresses for new sustainable designs.

Design/Methodology:

Forty-two test specimens were manufactured and analyzed according to the ASTM D695-15 standards. Eight numerical analyzes were performed on a real design manufactured with rPET using Young's compression modulus from the experimental tests. Finally, eight additional experimental tests under uniaxial compression loads were performed on the real sustainable design for validating its mechanical behavior versus computational numerical tests.

Findings:

As a result of the experimental tests, rPET behaves linearly until it reaches the elastic limit, along each manufacturing axis. The results confirmed the design's structural safety by the load scenario and operating boundary conditions. Experimental and numerical results show a difference of 0.001 mm to 0.024 mm, allowing for the rPET to be configured as isotropic in numerical simulation software without having to modify its material modeling equations.

Practical implications:

The results obtained are of great help to industry, designers, and researchers because they validate the use of recycled rPET for the ecological production of real-sustainable products using MEX technology under compressive stress and its configuration for numerical simulations. Major design companies are now using recycled plastic materials in their high-end designs.

Originality:

Validation results have been presented on test specimens and real ítems, comparing experimental material configuration values with numerical results. Specifically, no industrial or scientific work has been conducted with rPET subjected to uniaxial compression loads for characterizing experimentally and numerically the material using these results for validating a real case of a sustainable industrial product

## 1. Introduction

Applying sustainable design approaches that contribute to minimizing the environmental impact of products should assist in achieving the climate and environmental objectives required at the institutional level (Sadhukhan, J.. et al. 2021, Chatty, T. et al. 2022). Although these concepts included in the European directives (Regulation of the European Parliament 2022) seem a priori easy to apply; unfortunately, on a practical level, they cause many difficulties. Most decisions regarding the sustainability of a product are made during the design phase, so if rigorous data on the recycled material are not available in terms of functional and technological specifications, it is really difficult to carry out a virtual validation in the early stages of the development process. Some data, such as energy consumption, carbon footprint, recycled percentage, and material toxicity, have obvious ecological connections, however, it is not these but the mechanical, thermal, and chemical properties that play a more important role in minimizing the ecological impact (Jasti, N. et al. 2022, Merschak, S.et al. 2022).

The manufacture of the designed components transforms the raw materials into geometries suitable for the final assembly. The efficiency of the process is also decisive in the product's environmental impact (Gutowski, T. G et. al. 2009). In this line, additive manufacturing provides numerous advantages in terms of the sustainability of the manufacturing process. For example, the generation of less waste during manufacturing, the ability to optimize geometries, and the reduction of the consumption of materials and transportation in the supply chain (Ford, S. et.al 2016). Life cycle analyzes have shown that the use of additive manufacturing can produce significant savings in the production of goods. On the other hand, at a social level, a change in consumption patterns is taking place due to the ease of combining the use of TIC, CAD (Computer – Aided Design) software, and 3D printers. Users have gone from being completely passive consumers to producers of their designs becoming prosumers in a global manufacturing community.

Polymers, a fundamental material in additive manufacturing and injection molding, offer great opportunities for improvement in the field of sustainability (Mercado-Colmenero, J.M. et.al 2018,Wu, H. S. et.al 2021, Siddiqui, M. N et.al 2021). However, further research is required to explore and validate their mechanical properties for additive manufacturing to identify resource efficiency standards and allow standardization in the use of polymers. Waste plastic polymers can be reused and recovered in the form of recycled materials by transforming them into three-dimensional geometries using additive manufacturing (Collias, D. I. et al. 2021- Song J. et.al 2021, Pinter E. et al. 2021). Packaging is the largest market for components made from polymers, which mainly includes PE, PP, and PET materials. PET can be mechanically recycled into rPET granules or even blended with virgin PET to ensure higher quality. The interest in the rPET material is mainly motivated by the objective of printing from post-consumer waste such as food packaging (Hegyi, A et. al. 2022). Plastic polymers are characterized by their mechanical properties favorable to compressive stress (Celik, Y. et al. 2022- Chatham, C. A. et.al 2021). In this line, harnessing the advantages of polymers, especially those with a percentage of recycled material in the design of products subject to compressive stress can avoid the use of higher-cost materials such as metallic materials (Zhu. et. al. 2021, Nukala S. et al. 2022, Bartoli M. et al. 2022, Schneevogt, H. et. al. 2021).

Approaching the design of a new component safely implies validating its behavior using FEM numerical simulations (Mercado-Colmenero J.M et. al. 2020, Bustos Seibert M. et al 2022, Negoro et al. 2022). However, it is a fundamental requirement to previously know the mechanical properties of the material with which the component will be manufactured( Marciniak, D. et al 2018, Bączkowski M, et al 2021). For the specific case of recycled polymeric components manufactured using an additive deposition process, the mechanical properties differ enormously from the properties that the software possesses for a homogeneous material. This is not a trivial phenomenon since the manufactured part's mechanical performance depends on the geometry's complexity, the process's thermal conditions, and the gravitational loads (Tekoglu, C., et al. 2012- Morales N.G. et.al 2018). Unfortunately, this information is not available given that material data sheets only provide technical information related to the standard filament, not being valid to understand the mechanical behavior of the part in CAE analysis.

Numerical validation of an MEX (Material Extrusion) manufactured recycled polymer requires prior knowledge of its anisotropic material's mechanical behavior (Akhoundi, B. et. al. 2019, Garzon-Hernández,S. et al. 2020, Ngo, T.D. et al. 2018). To obtain these mechanical properties it is necessary to perform a series of experimental tests with specimens manufactured under the same conditions as the end piece (Sun, Q et al 2008). Finally, after obtaining the experimental information, it is necessary to find the best way to use it in the CAE software since the anisotropy and the low robustness of the manufacturing process make the usual finite element validation not feasible for these parts or assemblies (R.Lou et al. 2021, Domingo Espín, M. et. al. 2015). In this way, it is possible to obtain simulation results that will perfectly fit the experimental anisotropic behavior of the material. All these reasons make that designing and optimizing sustainable plastic parts manufactured by MEX continues to be a difficult engineering and research challenge (Ravari M.K. et. al. 2014). According to (Popescu D. et. al. 2018) and given the difficulty of fully characterizing the anisotropic behavior of materials manufactured using MEX technology, there are very few research works that present validation results on test specimens and real pieces at the same time, comparing, in turn, the experimental values of the material configuration with numerical results. Specifically, for the recycled rPET material subject to favorable uniaxial compression loads, there are no valid scientific or industrial data that characterize the material at an experimental and numerical level using these results for validating a real case of a sustainable industrial product.

To solve the proposed problems, the paper investigates the numerical and experimental modelling of the mechanical behaviour of the recycled polymer rPET additively manufactured by a sustainable MEX process and under gravitational compressive loads. Because of the anisotropy of the process and the variability in the quality of printed parts, finite element analysis is not directly applicable to recycled materials manufactured using fused filament fabrication. Specifically, no industrial or scientific work has been conducted with rPET subjected to uniaxial compression loads for characterizing experimentally and numerically the material using these results for validating a real case of a sustainable industrial product.

The results obtained are of great help to industry and researchers marking a turning point since it validates the use of the recycled polymer rPET for sustainable manufacturing of real sustainable design parts using a MEX manufacturing process as well as validates its mechanical behaviour for the numerical simulation of designs in terms of material and ecological production process. Major design companies are beginning to use recycled plastic materials in their high-end designs.

**1. Geometrical, functional, and MEX 3D manufacturing description of the design case study**

In this section, we proceed to describe the geometry, functionality, specifications, and MEX (Material Extrusion) process of the sustainable product. Likewise, mechanical analysis of the device is carried out to evaluate the stress state, load scenario, and boundary conditions to which it is subjected during its operation and service. The main objective is to determine how the different configurations of the 3D MEX process affect or intervene in its final structural behavior and performance for numerical uses. Fig. 1 shows the geometry of the design.

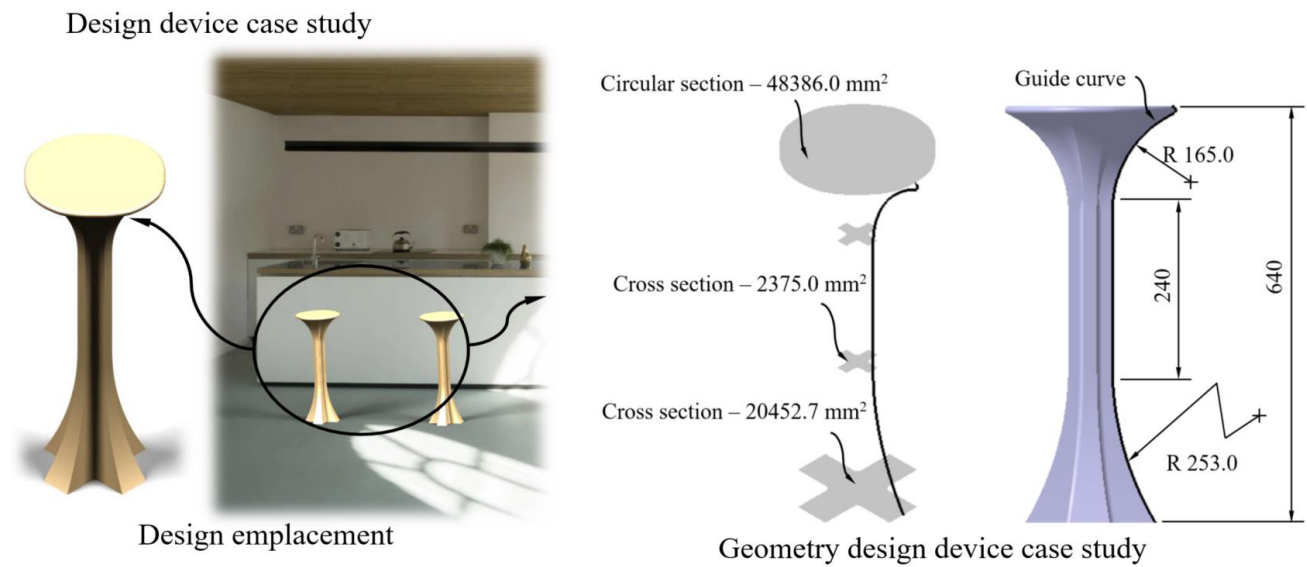

**Fig. 1** Design emplacement, geometry and graphic description of the design device manufactured with MEX

The functional element under study in the present manuscript is defined as a stool with application in domestic and urban environments. To meet the aesthetic requirements and get an organic shape, as shown in Fig. 1, this design stool has variable section geometry throughout its domain, which is 640 mm long. Fig. 1 and Table 1 show the magnitude and shape of the main geometric elements and parameters, at a 1:1 scale, to define the topology of the design stool.

Table. 1 Geometric parameters values of the design stool

| Variable | Units | Value |
|---|---|---|
| Base cross section area | mm$^2$ | 20452.7 |
| Mid cross section area | mm$^2$ | 2375.0 |
| Circular cross – section area | mm$^2$ | 48386.0 |
| Overall device height | mm | 640.0 |
| Top radius of guide curve | mm | 165.0 |
| Base radius of guide curve | mm | 253.0 |
| Straight length of guide curve | mm | 240.0 |

On the other hand, as shown in Fig. 3, the functionality and conditions of use and service of the design stool mean that its structural behavior is defined by a pure uniaxial compression stress state. That is, given that the applied load scenario is a unidirectional axial force applied at one stool end (see Fig. 2) and the boundary condition is a fixed support at the opposite end of it (see Fig. 2), the structural behavior of the geometry is pure uniaxial compression. As shown in Fig. 1, the geometric design of the stool is adapted to its functionality and its requirements of use. Its upper circular section (see Fig. 1) allows uniformly distributing the load towards the main column of the stool with a cross-section. Likewise, the section at the base of the stool gives it greater structural rigidity improving the stability and load transfer to the main support of the design stool.

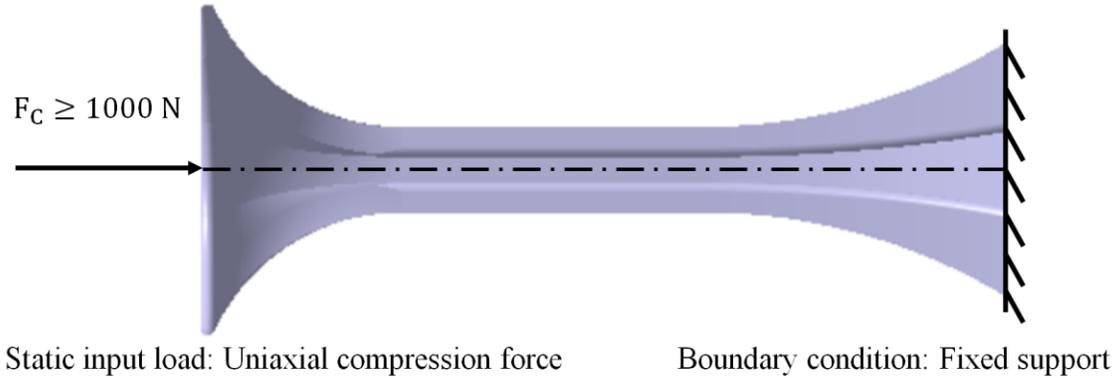

F$_C$ ≥ 1000 N

Static input load: Uniaxial compression force     Boundary condition: Fixed support

**Fig. 2** Boundary conditions and load definition for the design stool

In this way, the main objective is to carry out an experimental and numerical analysis of the structural behavior of the design and the characterization of the elastic and mechanical properties of the recycled plastic material rPET used for its manufacture. All this is for a tensional state of pure uniaxial compression. Thus, from the uniaxial compressive force applied to the end of the stool (see Fig. 2) and the tensile yield stress of the plastic material, the compressive stress field to which the design stool e is subjected is determined by a classical structural analysis (see Eq. 1).

$$\sigma_{ys} > \sigma_c = \int_{L_o}^{L} \frac{F_c}{A_{cross}(l)} \cdot dl \qquad (1)$$

Where $\sigma_{ys}$ [MPa] represents the yield stress of the material, $\sigma_c$ [MPa] represents the compression stress to which the design stool is subjected, $F_c$ [N] represents the uniaxial compression force (see Fig 2), and $A_{cross}$ [mm$^2$] represents the variable cross-section of the design stool along its length L [mm] (see Fig. 1). As shown in Fig. 2, according to the structural requirements to which the design stool is subjected, it must be able to withstand a service limit state in which the compression stress $\sigma_c$ [MPa] (see Eq. 1) should not exceed the elastic limit of the material, $\sigma_{ys}$ [MPa] (see Eq.1). Where the compression stress $\sigma_c$ [MPa] (see Eq. 1) is defined by the uniaxial compression force of 1000 N, applied on the variable section of the design stool (see Fig. 2).

After defining the structural and functional requirements, as well as the topology and geometric parameters of the design stool, these must be adapted to the main features of the experimental testing machines and mechanisms used to carry out the experimental validation of the operation and structural behavior of the stool. For this, it should be noted that a reduction scale of magnitude 1:4 is applied to all the dimensions of the design. In addition, as shown in Fig. 1, the end sections of the design stool have been defined to facilitate its fastening on the supports and clamps of the measuring instruments and of the uniaxial compression testing machine, keeping its geometry and cross-shaped central section constant for the main area of analysis of the case study.

In terms of manufacturing, the material used to make the design stool is rPET (recyclable polyethylene terephthalate) (3R3D Technology Materials 2022). The optimum extrusion temperature for this material is 255ºC, with 70ºC being the value of the 3D printer's heated bed recommended temperature. These thermal parameters facilitate the adhesion between layers, reducing defects caused by excess viscosity and reducing stress concentrations between layers. Table 2 and Table 3 show the magnitude of the physical and printing parameters of this material. All these features make this material suitable for structural designs located in interior spaces.

The design is manufactured using MEX technology, using the commercial printer model (Ultimaker 2+ Extended) (see Fig. 3). Its main dimensions are 23 mm on the X axis, 223 mm on the Y axis, and 305 mm on the Z axis. The thermal treatment and the conditions to which the material is subjected during the manufacturing process cause a variation in the mechanical and elastic properties concerning the properties of the original plastic filament. Moreover, the versatility of this manufacturing process allows the use of different printing paths and varying the building orientation of the part or the direction of deposition of layers of material. For this reason, to characterize the material after the manufacturing process and establish the printing path and direction that optimizes the mechanical resistance of the design stool, 42 cylindrical geometry test specimens have been defined. As shown in Fig. 3, given the geometry of revolution presented by the design stool and the test specimens, the Y/X axis, and Z axis are defined as the main printing directions. Furthermore, four strategies and trajectories for the deposition of the melt material have been defined: concentric, Zig – Zag, Zig – Zag with θ = – 45º/45º between layers and Zig – Zag θ = 0º/90º between layers. The geometric dimensions of the cylindrical specimens are established according to the applicable standard used for the characterization of material (ASTM D695 – 15), this being equivalent to the European standard (ISO 604 – 2003). In this way, the total height of the cylindrical specimen is 50.8 mm and the diameter of its circular section is 12.7 mm. The selection of this type of test specimen is justified by its geometry since it has greater similarity and shares geometric features with the geometry of the design stool, which allows establishing a characterization of the mechanical and elastic properties of the material closer to its conditions of use. On the other hand, to carry out the experimental validation of the design stool, 7 prototypes have been manufactured, maintaining the main printing directions, the deposition trajectories of molten material, and the printing parameters for the manufacture of the test pieces. As can be seen in Fig. 4, to carry out the 3D additive manufacturing of the design stool in the X/Y direction, the generation of supports is necessarily required. These supports are made with the same material, rPET, and are removed manually after the manufacturing process. The printing parameters of the supports differ from those established for the main geometry of the design stool. Table 2 shows the magnitude of the technological parameters used during the manufacturing process, for cylindrical specimens and prototypes of the design stool. The magnitudes of the printing parameters, shown in Table 2, have been established based on the recommendations of the supplier of the rPET material filament.

Ensuring that the parameters and configuration of the 3D additive manufacturing process used for the cylindrical test specimens are the same as those used for the design stool is crucial to obtaining reliable results and a complete characterization of the rPET plastic material. Any deviation from the original parameters and configuration used for the design stool could lead to significant changes in the mechanical and structural behavior of the material. Therefore, it is important to maintain the same parameters and configuration of the additive manufacturing process throughout the experimental tests to accurately analyze the behavior of the design stool and ensure that the results obtained are valid.

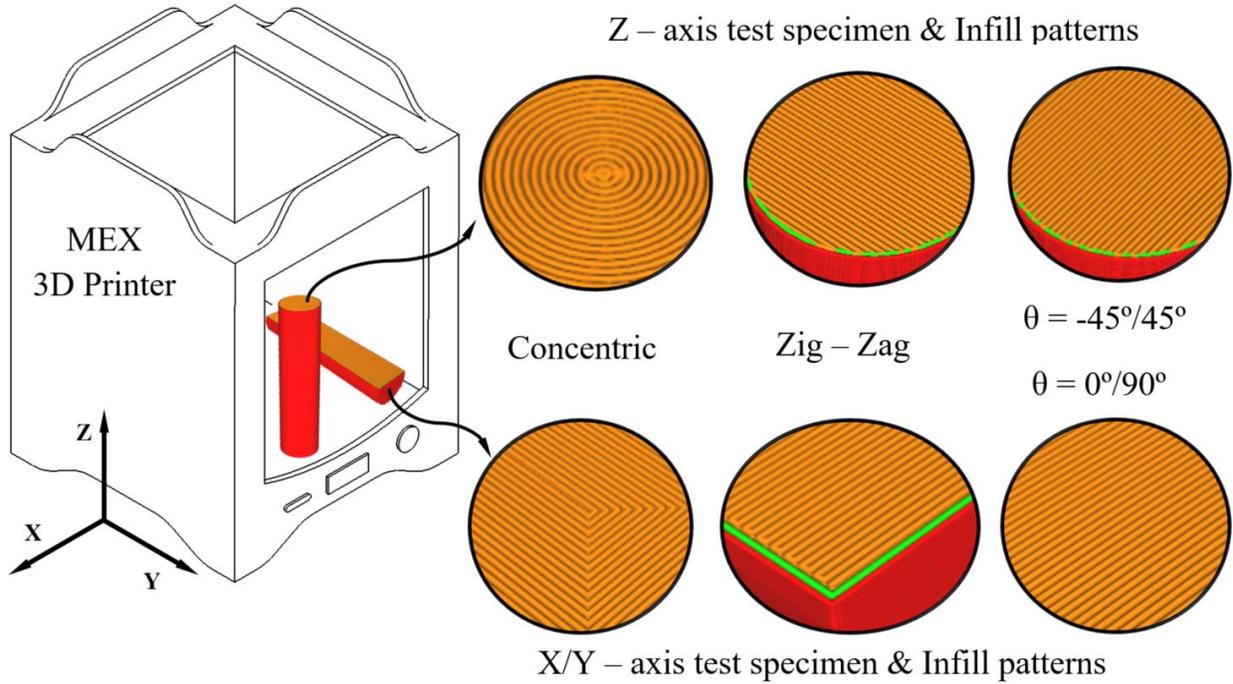

**Fig. 3** MEX process configuration for test specimen manufacturing

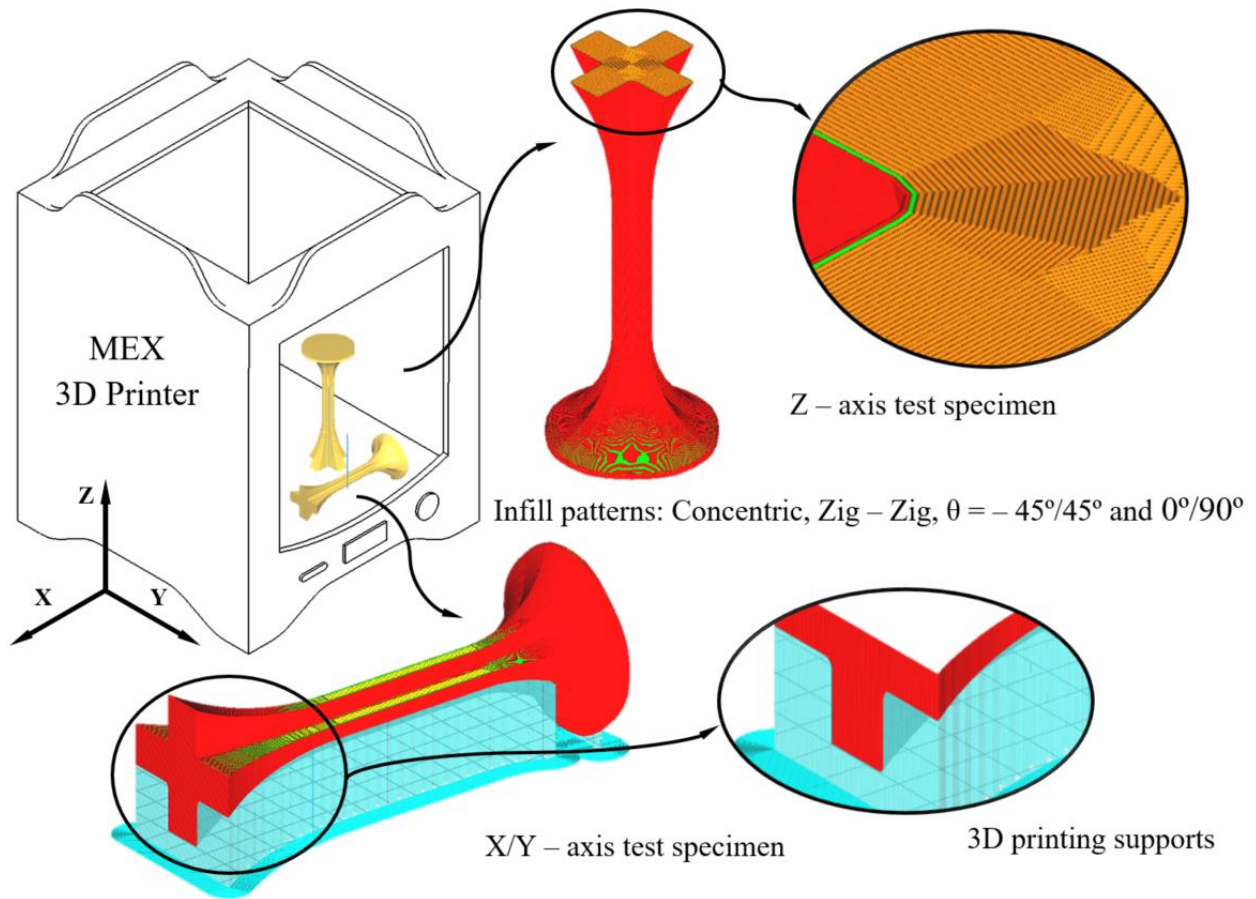

**Fig. 4** MEX process configuration for design device manufacturing

**Table. 2** Manufacturing parameters of the MEX process for the cylindrical specimens and design stool

| Manufacturing parameters | Units | Value |
|---|---|---|
| Layer height | mm | 0.15 |
| Line width | mm | 0.30 |
| Contour lines | mm | 0.30 |
| Wall thickness | mm | 0.60 |
| Wall count | – | 2 |
| Outer wall wipe distance | mm | 0.20 |
| Infill density | % | 100 |
| Infill line distance | mm | 0.20 |
| Infill line width | mm | 0.30 |
| Infill layer thickness | mm | 0.15 |
| Infill / Wall speed | mm/s | 38 |
| Nozzle diameter | mm | 0.40 |
| Extruder definition | – | Ultimaker Print Core AA - 0.4 mm |
| Infill pattern | – | Concentric, Zig – Zag, $\theta = -45°/45°$, $\theta = 0°/90°$ |
| Wall pattern | – | Concentric |
| Support pattern | – | Giroid |
| Support density | % | 20 |
| Support distance Z – X/Y | mm | 0.2 – 0.75 |
| Extrude temperature | ºC | 255 |
| Buildplate temperature | ºC | 70 |

## 2. Materials

### 2.1. Recyclable polyethylene terephthalate

The rPET plastic filament used in the present manuscript was obtained from the supplier 3R3D Technology Materials, s.l. (Irun, Spain) While the specific processing parameters used to manufacture the filament are confidential, the supplier has indicated that the material is manufactured in accordance with the UNE-CEN/TS 16861:2015 and UNE-EN 15348:2015 standards. Figure 5 shows the spools in which the filament was supplied by the manufacturer. The characterization of the properties of the material is conditioned by the manufacturing process to which it is subjected; in particular, the manufacturing process established for the characterization of the material is MEX technology. In this way, although the elastic, physical, and mechanical properties of the plastic filament are defined by the manufacturer and supplier of the material, these vary significantly during the manufacturing process being conditioned by the technological and geometric parameters that govern the process. Therefore, to perform a complete and exhaustive characterization of the material, it is necessary to establish and determine the elastic, physical, and mechanical properties of the material after undergoing the 3D additive manufacturing process. Likewise, it should be noted that the stress state to which the material is subjected directly influences the definition and dimensioning of the technological and geometric parameters that determine the 3D additive manufacturing process. For this reason, the characterization of the material is carried out according to the stress state to which it is subjected. In particular, the methodology developed assumes that the stress state to which the material is subjected is pure uniaxial compression. rPET is a recyclable plastic whose main composition is established by a percentage of pure PET (Polyethylene terephthalate), which ranges between 30%~50%, and a percentage of recycled PET, which ranges between 70%~50%. In particular, the rPET supplied by the manufacturer and supplier 3R3D Technology Materials, s.l. has a composition of 35% pure unrecycled PET and 65% high-quality recycled PET. Table 3 shows the elastic, physical and mechanical properties of the filament of pure PET and recycled rPET. The magnitude of the properties described in Table 3 have been provided by the manufacturer of the filament of the plastic materials, 3R3D Technology Materials, s.l.

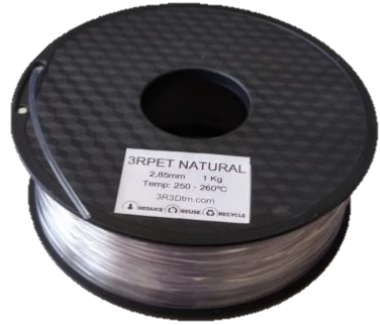

**Fig. 5** Spools of rPET plastic filaments supplied by the manufacturer 3R3D Technology Materials, s.l

**Table. 3** Mechanical and physical properties features for PET and rPET

| Properties | Units | Pure PET | Recycled PET |
|---|---|---|---|
| Young's modulus | MPa | 1890 | 1630 |
| Tensile yield strength | MPa | 47 | 24 |
| Elongation strength | % | 83 | 110 |
| Impact strength | J/m | 12 | 20 |
| Melting temperature | ºC | 244 – 254 | 247 – 253 |
| Density | g/cm$^3$ | 1.380 | 1.380 |
| Molecular mass | g/mol | 81,600 | 58,400 |

The treatment process, for the percentage of recycled PET that makes up the rPET, is generated through mechanical recycling. For the material, the initial waste comes from clean water bottles with the same composition and color. First of all, the waste undergoes an initial separation stage to eliminate possible impurities from other materials with which it has been in contact. In this case, as it is a waste from clean plastic bottles, the separation process is carried out manually. Next, the selected waste is subjected to a grinding process using mills or grinders to obtain a PET plastic flake smaller than 10 mm and free of dust. Next, the washing phase is carried out with water, surfactants, and diluted soda at a temperature of 70 ºC. At the end of this process, the surfactants and the diluted soda used are removed by successive washings with water at room temperature. Lastly, the cleaned flake is dried at 150 ºC and subjected to an extrusion process, at high temperature and pressure, to generate the final recycled PET. From the pellets generated and together with the contribution of pure PET, the filament of rPET is generated. This mechanical recycling process, together with the contribution of pure PET, allows the homogenization of the final rPET.

### 3. Experimental tests

To develop the objectives established, the research work has focused on the development of two types of experimental tests. In the first of them, the structural and mechanical behavior of the design stool has been analyzed, according to the pure uniaxial compression stress state to which it is subjected (see Fig. 2). On the other hand, in the second test, the structural and mechanical behavior of the different configurations of rPET test specimens manufactured using MEX technology has been experimentally analyzed. This type of test has been established following the guidelines defined in the international standard ASTM D695-15. Its objective is to characterize the mechanical and elastic properties of the rPET under a pure uniaxial compression stress state. In other words, the experimental results of this test determine elastic and mechanical properties such as Young's modulus, yield stress, and fracture stress, of the rPET after undergoing 3D additive manufacturing conditions. In addition, the mechanical and elastic characterization of the plastic material, based on the testing of specimens, allows establishing the properties and features necessary to carry out its definition and modeling in commercial numerical simulation software of the FEM (Finite Element Method). In this way, the methodology developed can be validated by comparing the results obtained from the numerical simulations with the results obtained from the experimental tests.

According to the international standard ASTM D695-15, at least 5 cylindrical specimens must be tested in each of the main directions of 3D additive manufacturing. To accomplish the international standard for each 3D additive manufacturing configuration, 6 cylindrical test specimens have been manufactured. Thus, since the three main 3D printing directions and four profile pathing are analyzed, a total of 42 cylindrical specimens have been mechanically tested during the development of the presented methodology. It should be noted that, given the geometric condition of revolution that the analyzed cylindrical specimens possess, the experimental results obtained for the main manufacturing direction Z in the printing patterns θ = – 45º/45º and θ = 0º/90º are analogous. Likewise, after the manufacture of each cylindrical specimen, a measurement has been made of them to check their dimensions and use them, later, for the calculation of the mechanical and elastic properties of the plastic material for each manufacturing configuration used. Table 4 shows the average value of the main dimensions obtained together with the magnitude of their dispersion.

**Table. 4** Mean dimensions and dispersion for the 3D printing cylindrical specimens

| Profile pathing | Concentric | | Zig - Zag | |
|---|---|---|---|---|
| | Diameter [mm] | Height [mm] | Diameter [mm] | Height [mm] |
| X/Y | 12.28±0.098 | 50.55±0.084 | 12.30±0.055 | 50.67±0.052 |
| Z | 12.53±0.052 | 50.37±0.052 | 12.45±0.066 | 50.36±0.066 |
| Profile pathing | θ = – 45º/45º | | θ = 0º/90º | |
| | Diameter [mm] | Height [mm] | Diameter [mm] | Height [mm] |
| X/Y | 12.29±0.038 | 50.68±0.042 | 12.47±0.041 | 51.05±0.055 |
| Z | 12.53±0.061 | 50.37±0.061 | 12.53±0.061 | 50.37±0.061 |

After completing the manufacturing process of each cylindrical test specimen, its experimental test is performed according to the guidelines established in the international standard ASTM D695-15. Following the international standard ASTM D695-15 (2015), the experimental uniaxial pure compression test is developed at an approximate speed equal to 1 mm/min (see Eq. 2) until structural failure occurs and the cylindrical specimens break.

$$V_c = 0.02 \cdot L_s \quad (2)$$

Where $V_c$ [mm/min] represents the compression speed of the clamps of the testing machine and $L_s$ [mm] represents the total height of the cylindrical specimens analyzed (see Table 4). Likewise, as can be verified, said compression speed of the machine jaws, in each of the experimental tests carried out, complies with the requirements defined in the international standard ASTM D695-15. The compression speed of the machine jaws must be equal to 1.3 ± 0.3 mm/min so that the compression machine MTS – 810 reaches the requirements established in the ASTM D695-15 standard. Moreover, to complete the description of the testing machine characteristic, the magnitude of the force capacity is 500 kN, the vertical test dimension is 2108 mm and its stiffness is equal to $7.5 \cdot 10^8$ N/m.

Fig. 6, Fig. 7, Fig. 8, and Fig. 9 show the results obtained from the uniaxial pure compression experimental tests for the main directions, X/Y and Z, of additive 3D printing and for each of the trajectories of 3D additive printing, concentric, Zig – Zag, θ = – 45º/45º and θ = 0º/90º, used, respectively, for the manufacture of the cylindrical specimens analyzed. Moreover, Fig. 6, Fig. 7, Fig. 8, and Fig. 9 show the evolution of the pure uniaxial compression force and nominal displacements to which each of the cylindrical specimens analyzed during the experimental tests are subjected.

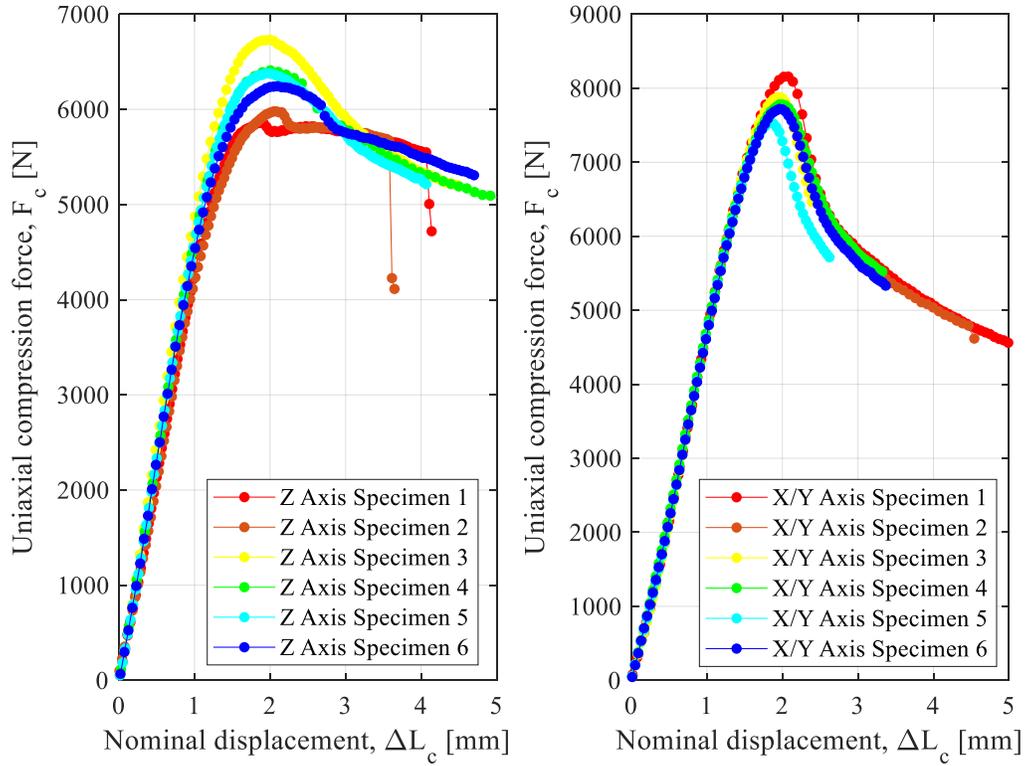

**Fig. 6** Curves for concentric pathing – uniaxial compression force versus nominal displacement

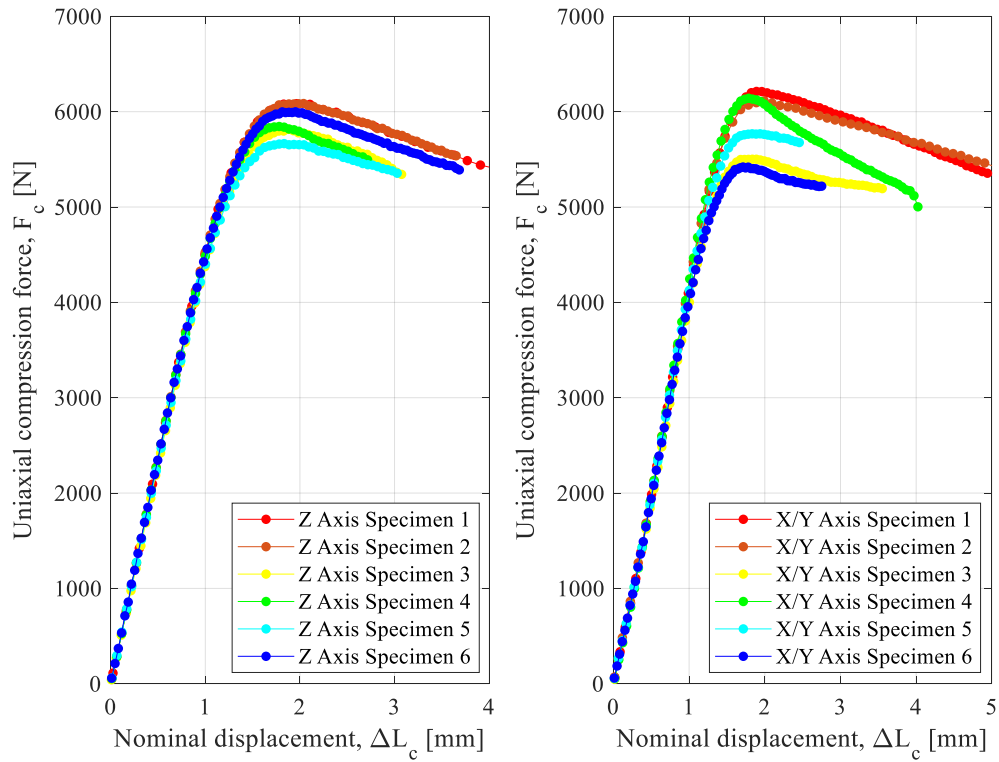

**Fig. 7** Curves for Zig – Zag pathing – uniaxial compression force versus nominal displacement

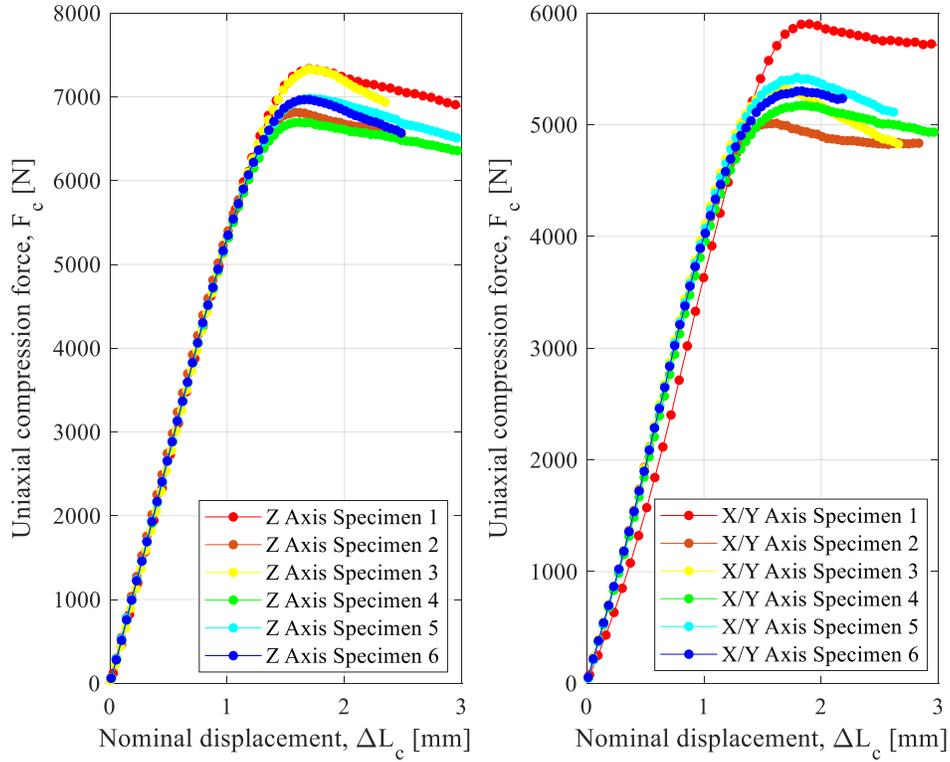

**Fig. 8** Curves for θ = − 45º/45º pathing – uniaxial compression stress versus nominal strain

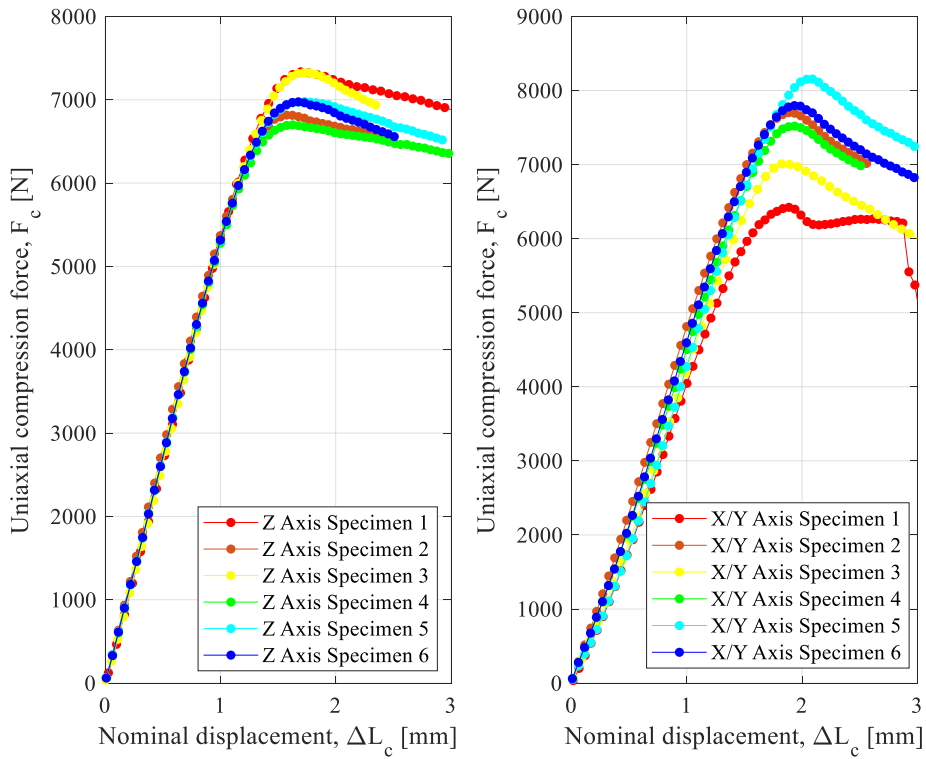

**Fig. 9** Curves for θ = 0º/90º pathing – uniaxial compression stress versus nominal strain

As shown in Fig. 6, Fig. 7, Fig. 8, and Fig. 9, the data obtained from the set of experimental tests performed cover from the beginning of the test, until the moment of structural failure or breakage of each cylindrical test specimen analyzed. Furthermore, the results of the uniaxial compression force and the nominal displacement field obtained from the measuring instruments of the testing machine, it is possible to determine both the magnitude of the uniaxial compression stress field and the field of nominal strains. All this is according to the equations and methodology described in the international standard ASTM D695-15 (see Eq. 3 and Eq. 4).

$$\sigma_c = \frac{F_c}{A_{cross}} \qquad (3)$$

$$\varepsilon_c = \frac{\Delta L}{L_s} \qquad (4)$$

Where $\sigma_c$ [MPa] represents the compression stress field to which the cylindrical specimens are subjected during the test, $F_c$ [N] represents the compression force to which the cylindrical specimens are subjected during the test, $A_{cross}$ [mm$^2$] represents the circular sectional area of the cylindrical specimens (see Table 4), $\varepsilon_c$ [mm/mm] represents the nominal deformation to which the cylindrical specimens are subjected during the test, $\Delta L$ [mm] represents the nominal displacement experienced by the cylindrical specimens during the test and $L_s$ [mm] represents the initial length of the cylindrical specimens (see Table 4). From the uniaxial pure compressive stress field and the resulting nominal strain field, it can be determined that the rPET has a hardening elastic behavior from the beginning of its deformation until reaching the yield strength at a compression value $\sigma_{yc}$ [MPa]. From the yield stress magnitude, $\sigma_{yc}$, the softening plasticizing process of the material plastic rPET begins. This parameter determines the maximum magnitude of stress to which the plastic material rPET is subjected during the pure uniaxial compression tests and establishes the beginning of its fracture and structural collapse process. During this stage, the plastic material undergoes softening non–linear plastic deformation until reaching the fracture stress σfc [MPa], from which the cylindrical test specimens reach final structural failure. Table 5, Table 6, Table 7, and Table 8, show the magnitude of the arithmetic average and standard deviation of the main elastic and mechanical properties obtained after the experimental tests for each of the main manufacturing directions and 3D printing patterns used in the manufacture of the different cylindrical specimens analyzed. Also, according to the guidelines established in the international standard ASTM D695-15 (see Eq. 5 and Fig. 10), the compression Young's modulus is determined by the ratio of nominal stresses and strains between the initial and final value of the elastic and linear region of the curve $\sigma_c - \varepsilon_c$. Figure 10 and Eq. 5 are established by the ASTM D695-15 standard, which provides specific conditions, requirements, and guidelines for determining the mechanical and elastic properties of plastic materials under pure uniaxial compression load state.

$$E_c = \frac{\sigma_{cb} - \sigma_{ca}}{\varepsilon_{cb} - \varepsilon_{ca}} \qquad (5)$$

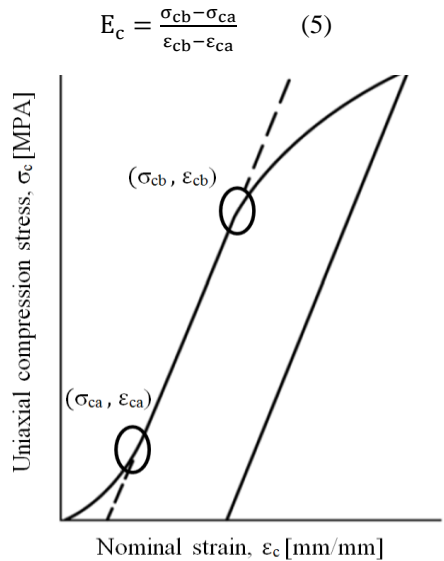

Fig. 10 Definition of the stress and strain values for Young's modulus calculation of the rPET

Table. 5 Elastic and mechanical compression properties for rPET concentric pathing specimens

| Mechanical properties | Units | X/Y Axis | | Z Axis | |
|---|---|---|---|---|---|
| | | Arithmetic average | Standard deviation | Arithmetic average | Standard deviation |
| Young's modulus, $E_c$ | [MPa] | 1,743.50 | 16.78 | 1,781.20 | 118.85 |
| Yield stress, $\sigma_{yc}$ | [MPa] | 61.65 | 1.51 | 49.51 | 2.26 |
| Fracture stress, $\sigma_{fc}$ | [MPa] | 42.21 | 5.36 | 41.99 | 1.69 |
| Yield strain, $\varepsilon_{yc}$ | [mm/mm] | 0.038 | 0.001 | 0.039 | 0.002 |
| Fracture strain, $\varepsilon_{fc}$ | [mm/mm] | 0.076 | 0.033 | 0.097 | 0.026 |

Table. 6 Elastic and mechanical compression properties for rPET Zig – Zag pathing specimens

| Mechanical properties | Units | X/Y Axis | | Z Axis | |
|---|---|---|---|---|---|
| | | Arithmetic average | Standard deviation | Arithmetic average | Standard deviation |
| Young's modulus, $E_c$ | [MPa] | 1,543.10 | 21.04 | 1,849.30 | 21.06 |
| Yield stress, $\sigma_{yc}$ | [MPa] | 33.96 | 2.09 | 27.92 | 1.18 |
| Fracture stress, $\sigma_{fc}$ | [MPa] | 39.91 | 2.97 | 42.61 | 0.65 |
| Yield strain, $\varepsilon_{yc}$ | [mm/mm] | 0.035 | 0.002 | 0.036 | 0.001 |
| Fracture strain, $\varepsilon_{fc}$ | [mm/mm] | 0.095 | 0.051 | 0.085 | 0.052 |

Table. 7 Elastic and mechanical compression properties for rPET $\theta = -45º/45º$ pathing specimens

| Mechanical properties | Units | X/Y Axis | | Z Axis | |
|---|---|---|---|---|---|
| | | Arithmetic average | Standard deviation | Arithmetic average | Standard deviation |
| Young's modulus, $E_c$ | [MPa] | 1,548.50 | 40.51 | 2,088.7 | 94.08 |
| Yield stress, $\sigma_{yc}$ | [MPa] | 31.79 | 3.22 | 37.44 | 1.94 |
| Fracture stress, $\sigma_{fc}$ | [MPa] | 38.69 | 1.86 | 51.52 | 1.64 |
| Yield strain, $\varepsilon_{yc}$ | [mm/mm] | 0.035 | 0.003 | 0.033 | 0.001 |
| Fracture strain, $\varepsilon_{fc}$ | [mm/mm] | 0.077 | 0.060 | 0.075 | 0.054 |

Table. 8 Elastic and mechanical compression properties for rPET $\theta = 0º/90º$ pathing specimens

| Mechanical properties | Units | X/Y Axis | | Z Axis | |
|---|---|---|---|---|---|
| | | Arithmetic average | Standard deviation | Arithmetic average | Standard deviation |
| Young's modulus, $E_c$ | [MPa] | 1,687.80 | 95.43 | 2,088.7 | 94.08 |
| Yield stress, $\sigma_{yc}$ | [MPa] | 37.30 | 4.07 | 37.44 | 1.94 |
| Fracture stress, $\sigma_{fc}$ | [MPa] | 44.72 | 3.72 | 51.52 | 1.64 |
| Yield strain, $\varepsilon_{yc}$ | [mm/mm] | 0.034 | 0.001 | 0.033 | 0.001 |
| Fracture strain, $\varepsilon_{fc}$ | [mm/mm] | 0.058 | 0.012 | 0.047 | 0.014 |

From the experimental results obtained, it is verified that, in general, and for all the 3D printing patterns analyzed, the manufacturing direction that presents the best elastic behavior is the Z direction. Well, as shown in Table 5, Table 6, Table 7, and Table 8, in this main printing direction the magnitude of Young's modulus parameter is greater than that obtained for the X/Y main printing direction. The main reason that justifies this elastic behavior of the rPET plastic material is that both the main direction of 3D printing and the direction of application of the uniaxial compression force are aligned. That is, in the main Z direction, the direction of deposition of layers of the plastic material during the manufacturing process is analogous to the direction in which the uniaxial compression force is applied during the experimental tests. While, in the main X/Y direction, the direction of deposition of plastic material layers during the manufacturing process is perpendicular to the direction in which the uniaxial compression force is applied during the experimental tests. This difference causes a greater structural rigidity in the cylindrical specimens analyzed and an improvement in the elastic properties of the rPET.

Furthermore, based on the experimental results obtained for the 3D printing patterns considered, it is determined that the pattern θ = 0º/90º is the configuration that presents the best elastic and structural behavior. Together with the two 3D printing directions analyzed, this printing pattern presents a greater magnitude of Young's modulus parameter compared to the rest of the printing patterns. The θ = 0º/90º pattern, in which layers of material are deposited in two orthogonal directions, exhibits superior mechanical properties in both directions. This configuration of additive manufacturing enhances the stiffness of the rPET plastic material, making it more resistant to deformation compared to other patterns. Finally, it should be noted that, given the geometric condition of revolution that the analyzed cylindrical specimens possess, the experimental results obtained for the main manufacturing direction Z in the printing patterns θ = – 45º/45º and θ = 0º/90º are analogous. Well, in this case, the structural behavior of both 3D printing configurations is equivalent.

Regarding the structural collapse of the plastic material and the fracture of the cylindrical specimens tested, it has been found that this process is directly associated with the main direction of 3D printing. On the one hand, for the specimens manufactured in the main Z direction, since the load application direction is analogous to the layer deposition direction, these are compressed until the plastic material reaches the yield stress, $\sigma_{yc}$ (see Table 5, Table 6, Table 7, and Table 8). At this moment, the plastic deformation of the material begins, generating the geometry of the cylindrical specimens to buckle and bend in its central cross area (see Fig. 11) and causing an eccentricity in the applied uniaxial compression force and, therefore, the appearance of a bending and shear component in the stress field. Next, the fracture process continues with delamination between adjacent layers of plastic material and brittle fracture of polymer filaments in the central area of the cylindrical test specimens. Although the fracture of the filaments of plastic material is generated brittly, the fracture of the cylindrical specimen as a whole is determined to be ductile, as shown in the Figure 6, Figure 7, Figure 8 and Figure 9 of the experimental tests. As it is shown in Table 5, Table 6, Table 7, and Table 8, the magnitude of the yield stress, $\sigma_{yc}$, and fracture stress, $\sigma_{fc}$, are similar, however, the nominal strains for these parameters are not. Therefore, the cylindrical specimens present a process of softening plastic deformation, before their final fracture, which is established by compression between layers of plastic material until they collapse. On the other hand, for the main X/Y printing direction, the cylindrical specimens present a ductile fracture establish by the structural collapse of the plastic filaments in the outer central regions of the cylindrical specimens (see Fig 11). Thus, when the cylindrical specimens reach the yield stress, $\sigma_{yc}$, brittle fracture of the plastic filaments occurs and the cohesive force that exists between adjacent layers of plastic material is overcome and they begin to delaminate, drastically reducing their mechanical strength until they reach total breakage. In conclusion, it can be determined that the fracture typology for the cylindrical specimens manufactured in the Z direction is ductile. Since, the magnitude of the yield stress, $\sigma_{yc}$, and fracture stress, $\sigma_{fc}$, are similar and the cylindrical specimens reach a greater plastic deformation over time before breaking, without reducing their mechanical resistance. Analogous, the fracture typology for the cylindrical specimens manufactured in the X/Y direction is ductile. Since, the cylindrical specimens do not allow hardening by plastic deformation and reduce their mechanical resistance drastically from the yield stress, $\sigma_{yc}$. In both cases, the structural collapse of the plastic material takes place in the central cross sections of the cylindrical specimens (see Fig. 11).

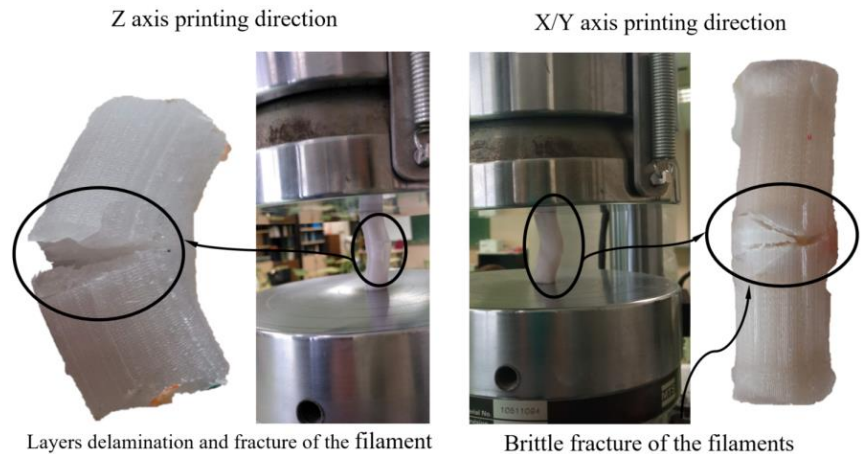

**Fig. 11** Structural collapse process of the cylindrical specimens during the experimental tests

Once the mechanical and elastic characterization of the rPET has been completed, the experimental tests of the design stool have been carried out. These experimental tests allow analyzing the real mechanical and structural behavior of the design stool for the load scenario and the boundary conditions to which it is subjected (see Fig. 2). The configurations defined for the 3D additive manufacturing process are analogous to those defined for the cylindrical test specimens. That is, 8 different prototypes of the geometry have been manufactured for the main X/Y and Z printing directions, and for the concentric, Zig – Zag, θ = – 45º/45º and θ = 0º/90º fill patterns (see Fig. 3). In this way, it is possible to compare and evaluate the influence that each configuration has on the mechanical behavior of the design geometry, as well as to determine the optimal configuration that maximizes its structural rigidity. To perform these experimental uniaxial pure compression tests, the MTS – 810 testing machine has been used again, defining a constant compression speed equal to 1 mm/min. That is, respecting the boundary conditions used for the mechanical and elastic characterization of the rPET. In this way, Fig. 12 shows the graph that relates the pure uniaxial compression force to which the different prototypes analyzed are subjected to the resulting displacement field, from the beginning of the experimental test to the structural collapse of the plastic material and its subsequent fracture.

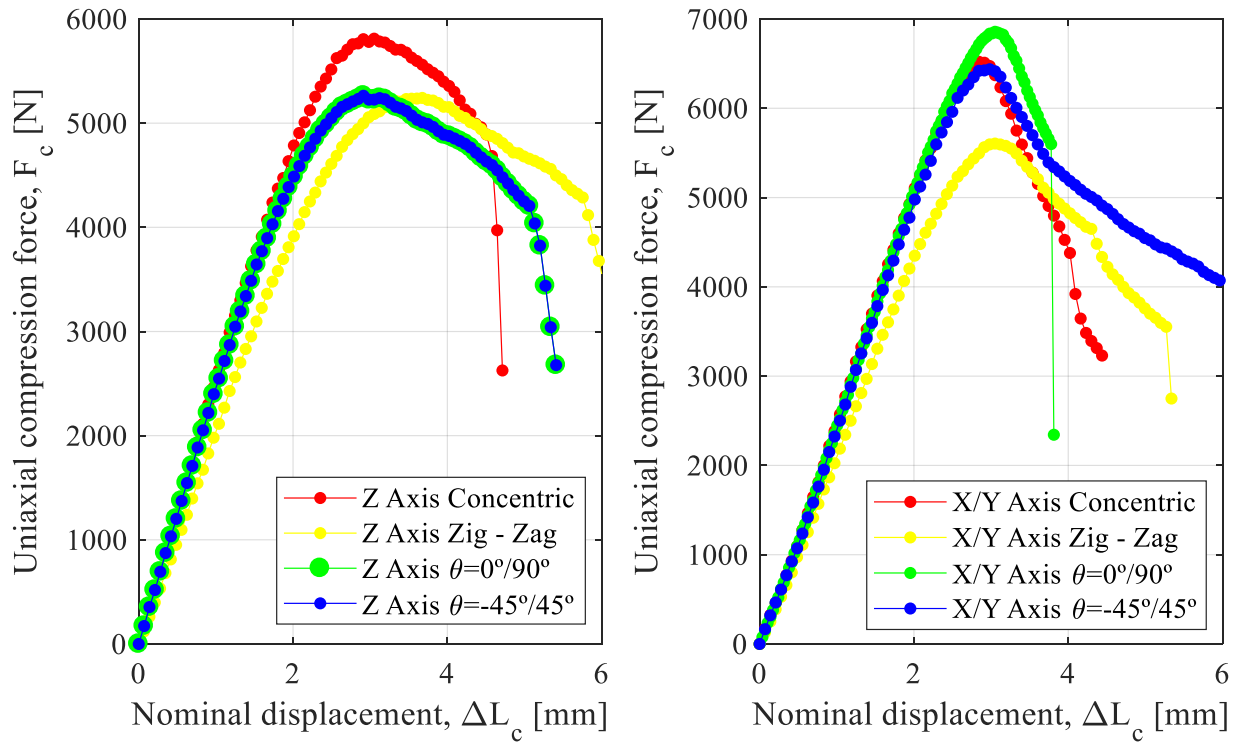

Fig. 12 Uniaxial compression force - nominal displacements for Z and X/Y axis prototypes

Table. 9 Mean dimensions and dispersion for the 3D printing prototypes tested

| Profile pathing | Concentric | | | | Zig - Zag | | | |
|---|---|---|---|---|---|---|---|---|
| | $F_{max}$ [N] | $\Delta L_{Fmax}$ [mm] | $F_{fc}$ [N] | $\Delta L_{fc}$ [mm] | $F_{max}$ [N] | $\Delta L_{Fmax}$ [mm] | $F_{fc}$ [N] | $\Delta L_{fc}$ [mm] |
| X/Y | 6,532 | 2.896 | 3,166 | 4.481 | 5,615 | 3.068 | 3,507 | 5.330 |
| Z | 5,813 | 2.955 | 4,830 | 4.551 | 5,243 | 3.640 | 4,287 | 5.755 |
| Profile pathing | $\theta = – 45º/45º$ | | | | $\theta = 0º/90º$ | | | |
| | $F_{max}$ [N] | $\Delta L_{Fmax}$ [mm] | $F_{fc}$ [N] | $\Delta L_{fc}$ [mm] | $F_{max}$ [N] | $\Delta L_{Fmax}$ [mm] | $F_{fc}$ [N] | $\Delta L_{fc}$ [mm] |
| X/Y | 6,440 | 2.956 | 3,673 | 6.890 | 6,869 | 3.068 | 5,533 | 3.809 |
| Z | 5,271 | 2.957 | 4,177 | 5.09 | 5,263 | 2.913 | 4,206 | 5.063 |

As shown in Table 9, $F_{max}$ [N] represents the magnitude of the maximum pure uniaxial compression force, $\Delta L_{Fmax}$ [mm] represents the nominal displacement for the maximum pure uniaxial compression force, $F_{fc}$ [N] represents the magnitude of the force for the fracture instant and ΔL_fc [mm] represents the nominal displacement for the fracture instant. From the experimental results obtained, as shown in Fig. 12 and Table 9, it can be determined that the elastic and mechanical behavior, as well as the plastic deformation and fracture process of the tested prototypes, differ according to their main manufacturing direction. On the one hand, the prototypes manufactured in the X/Y direction reach a higher magnitude of maximum uniaxial compression force, and, once this is reached, the process of fracture and structural collapse of the plastic material begins rapidly. In other words, these prototypes present a type of brittle fracture and, therefore, a lower hardening due to the plastic material plasticizing.

On the other hand, the prototypes manufactured in the Z direction present a lower maximum uniaxial compression force, but their process of fracture and structural collapse of the plastic material is more dilated in time. In this way, once the maximum uniaxial compression force is reached, the prototypes begin to soften due to the plastic deformation of the material, which admits a magnitude of larger nominal displacements. That is, these prototypes present a typology of ductile fracture. Moreover, as far as the fill patterns are concerned, the configuration with the greatest structural rigidity is the fill pattern θ = 0º/90º. Well, analogously to the elastic and structural behavior of the tested specimens, this configuration improves, as a whole, the magnitude of the parameters obtained after the experimental tests concerning the rest of the filling patterns. Finally, it should be noted that for each of the configurations defined for the manufacture of the prototypes by 3D printing, the maximum force obtained exceeds the magnitude of the service limit conditions defined for the design (see Fig. 2). That is to say, despite carrying out the experimental tests with prototypes subject to a reduced scale, these comply with the established mechanical requirements and, therefore, the full-scale design will also comply with them. Ensuring that the parameters and configuration of the additive manufacturing process are maintained and respected is crucial when applying the mechanical and elastic characterization of rPET plastic material obtained from cylindrical test specimens to a full-scale design. Any variation in the 3D printing parameters can affect the mechanical behavior of the material, and therefore, impact the overall performance of the design. It is important to replicate the same set up and configuration of the 3D printing process used for the cylindrical test specimens to ensure that the mechanical and elastic properties are consistent with the full-scale design. This will allow for accurate predictions of the performance of the full-scale design and ensure that it meets the established mechanical requirements.

## 4. Fractography

To evaluate the process of fracture and structural collapse of the recycled rPET, used for the different 3D additive manufacturing configurations, the fractology of the cylindrical specimens tested has been analyzed using a high-resolution scanning electron microscope (FESEM). The model of the electron microscope used is Carl Zeiss: Merlin. To evaluate the fracture that is generated in the cylindrical test specimens analyzed, four representative samples have been selected for the two main printing directions defined for their 3D additive manufacturing.

Fig. 13 and Fig.14 show with different details to a precision of 500 μm and 100 μm that the fracture mechanics that develops the cylindrical specimens manufactured in the main Z direction with a Zig – Zag and concentric patterns is ductile type and is mainly generated by the separation into sheets of adjacent layers of extruded rPET yarns. Fracture of the extruded yarns in both transverse and longitudinal directions divides the original layers into several layers that remain adhered to each other after breaking. The fracture evolution begins with a buckling process that weakens the circular layers of the specimens, generating both shear and bending forces that influence extruded yarns from contiguous sections during the layer separation process. Fig. 13 and Fig. 14 demonstrate how the longitudinal breakage of the extruded yarn generates thin and long yarns on the surface of the detached layers. This plastic fracturing process creates long strands of plastic material that break apart and become partially attached to the divided layers. Moreover, when delamination between adjacent layers begins due to molecular diffusion or overcoming interlayer cohesion force, a brittle fracture occurs in the plastic filaments close to the delaminated layers. The shear forces, derived from the bending of the cylindrical specimens during their plasticizing process, cause the plastic filaments to fracture transversely and generate crack growth in the longitudinal direction of the specimens. It is worth noting that both fractography and the crack growth process for the Z axis printing direction are similar for the different filling patterns employed in the additive manufacturing process.

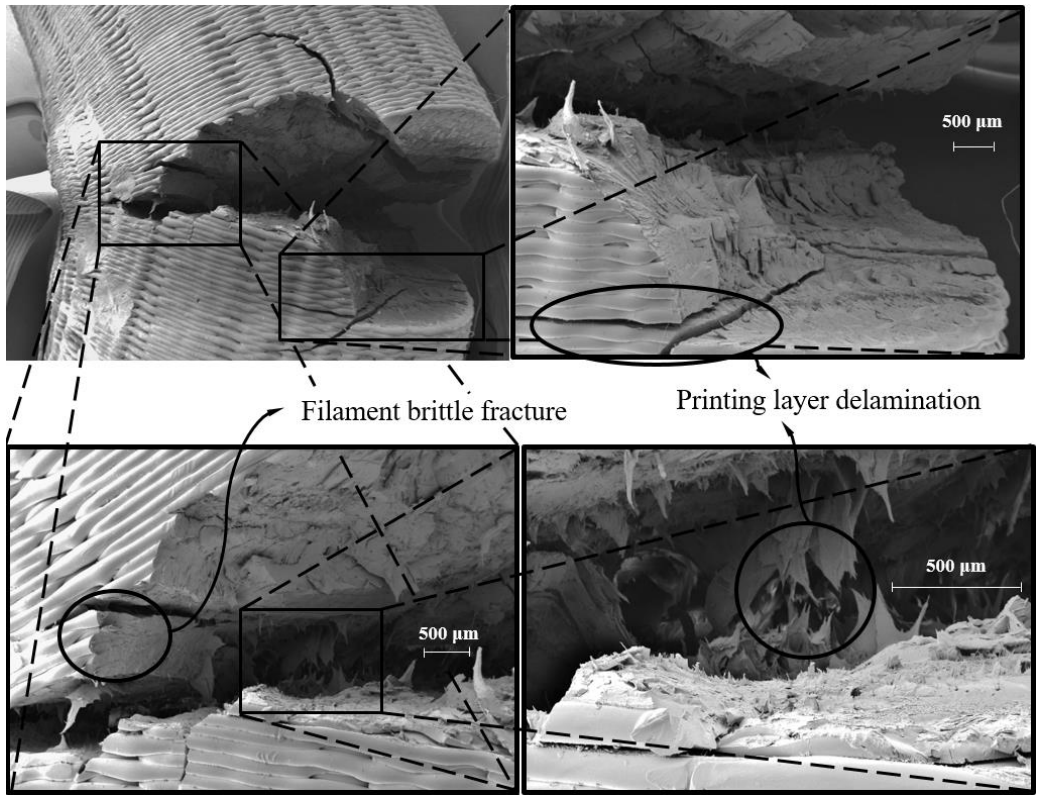

**Fig. 13** Cross section FESEM fracture images of Zig – Zag pathing and Z axis printing direction specimens

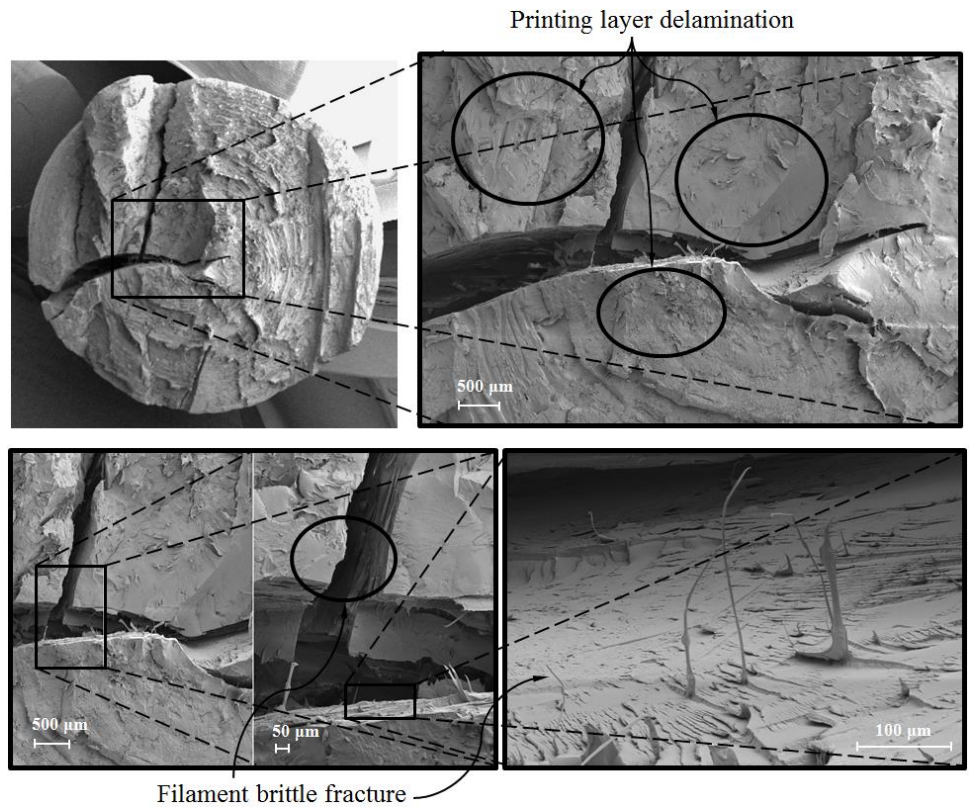

**Fig. 14** Cross section FESEM fracture images of concentric pathing and Z axis printing direction specimens

On the other hand, Fig. 15 and Fig. 16, with different details at a precision of 1 mm, 500 μm, 100 μm and 50 μm, shows the fracture mechanics and the resulting permanent plastic deformation in the cylindrical specimens manufactured in the main X/Y direction with Zig – Zag and concentric pattern, respectively, after the performed the experimental tests. As shown in Fig. 15 and Fig. 16, the fracture mechanics obtained for this type of specimen is caused by a buckling process that generates bending and shear stresses derived from the rotation and eccentricity suffered by the specimens when they are plastically deformed. That is, once the plastic material reaches the yield stress, $\sigma_{yc}$, it begins to deform plastically in the central sections of the specimens, causing an eccentricity in the applied compression force which causes bending and shear stresses that weaken the outer layers of the specimens. Then, after crack growth of the plastic filaments begins in the central sections of the cylindrical specimens, the shear stress field causes the molecular diffusivity or the interlayer cohesive force to be exceeded. That is, once the brittle and transverse fracture of the plastic filaments has occurred, the delamination process begins between adjacent layers of the cylindrical specimens oriented in the main X/Y direction, especially for their central layers. The behavior of the cylindrical specimens printed in the X/Y direction was different from those printed in the Z direction. In the former, a complete and clean break of the cross section did not occur, unlike the latter where a clear delamination process was observed. This is due to the fact that the specimens in the X/Y direction present greater plastic deformation, which causes the fracture process of the longitudinal plastic filaments in the central sections of the specimens to complete. However, Fig. 15 and Fig. 16 accurately show the brittle fracture process of the rPET plastic filament during structural failure of the printed specimens in the X/Y direction. It should be noted that both fractography and the crack growth process are analogous for the different filling patterns defined in the additive manufacturing process.

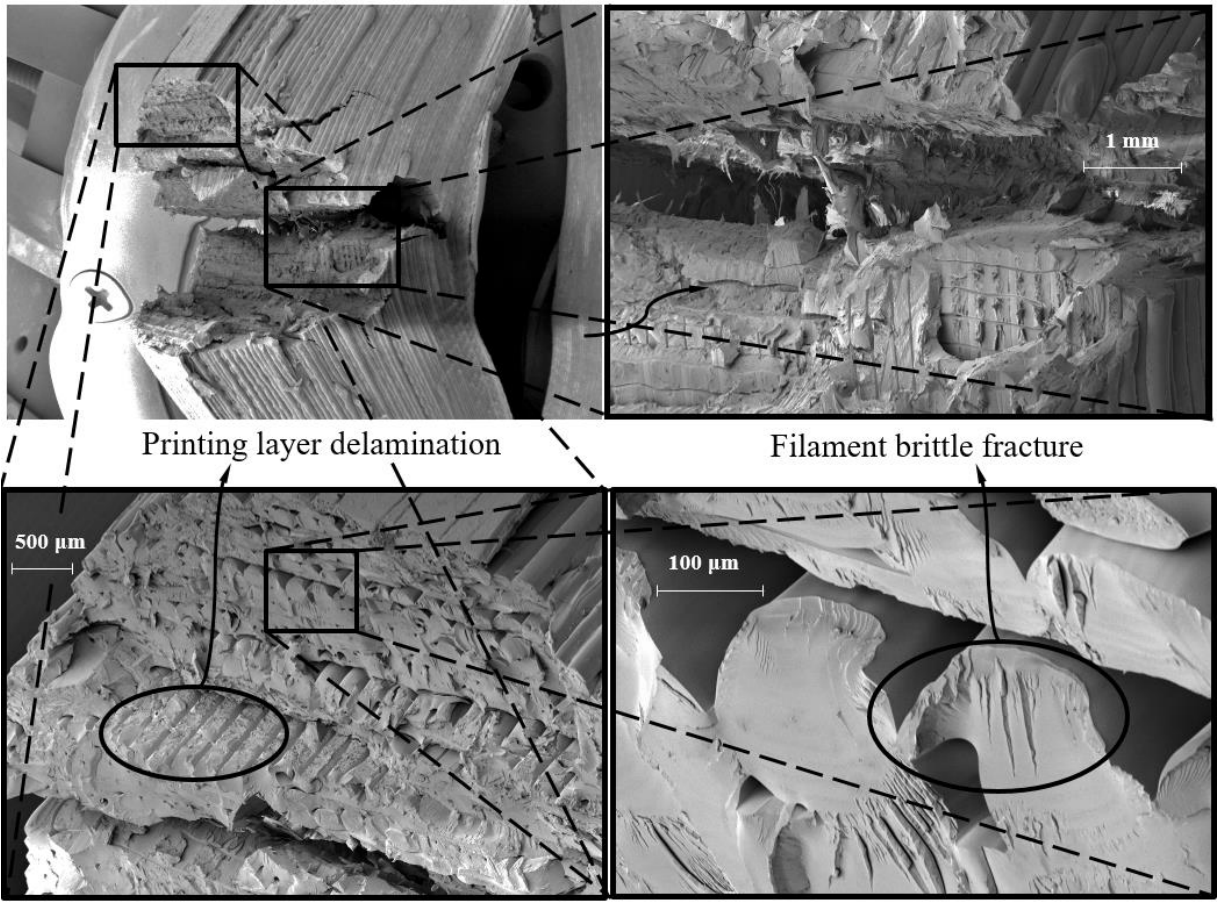

**Fig. 15** FESEM fracture images of Zig – Zag pathing and X/Y axis printing direction specimens

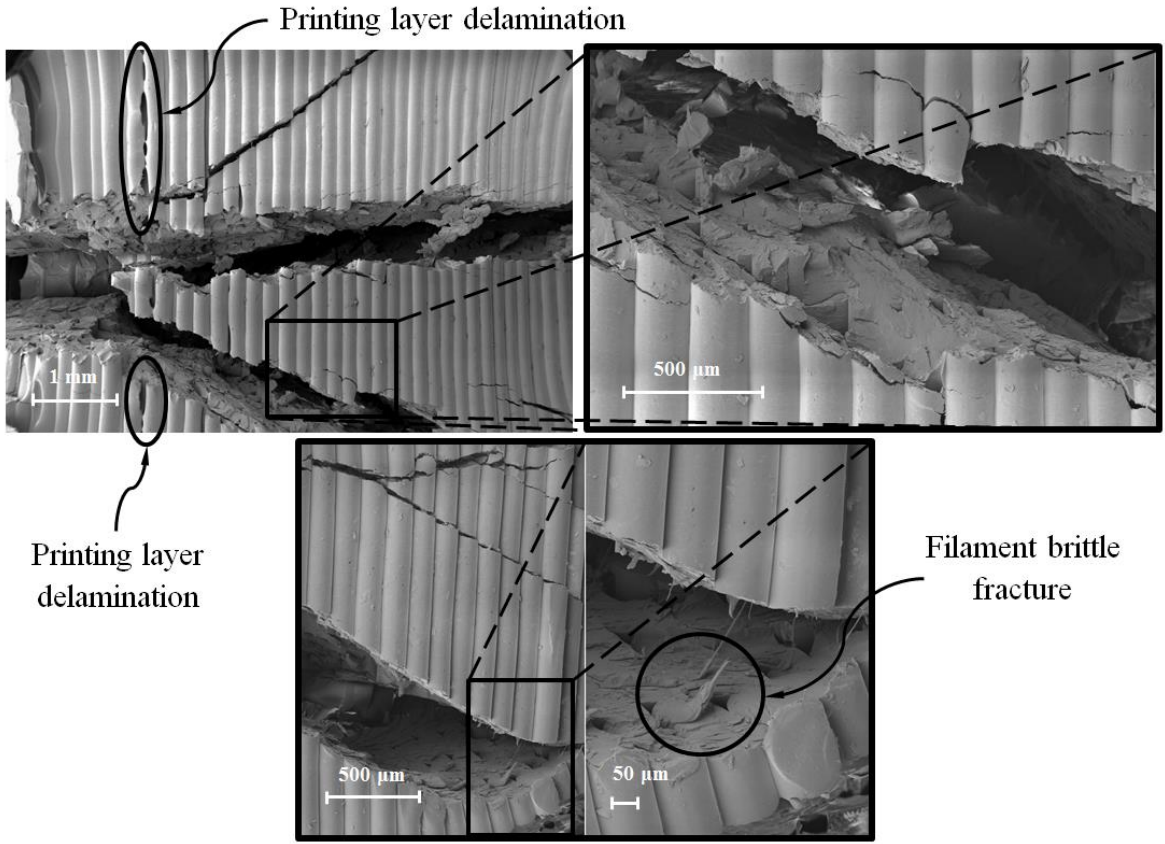

**Fig. 16** FESEM fracture images of concentric pathing and X/Y axis printing direction specimens

## 5. Numerical method

To perform the simulations of the designed stool, the FEM Ansys Mechanical 16.0 software has been used Fig. 2 and Fig. 17 indicate the boundary conditions and loads used for the structural validation of the geometry. The design load scenario focuses on a compression force aligned with the longitudinal axis of the applied geometry in the upper support area of magnitude 1000 N. Regarding the boundary conditions, a fixed support has been established in the lower area or base of the geometry (see Fig. 17).

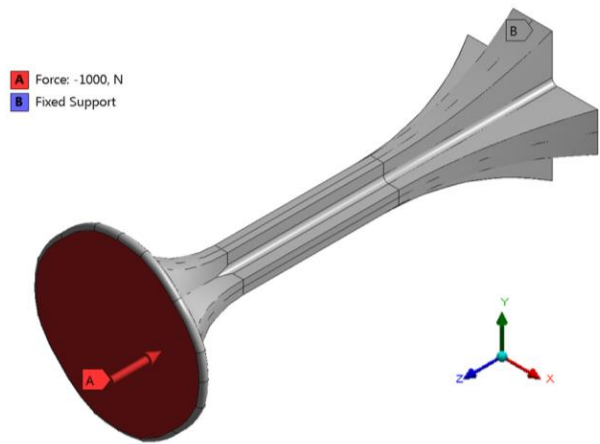

**Fig. 17** Boundary conditions and loads scenario for the numerical simulation of the design stool

The definition of the numerical model used to perform the mechanical analysis has been elastic and linear. Eight numerical simulations have been performed to analyze the structural behavior of the geometry for each of the different 3D additive manufacturing configurations described (see Fig. 4). The main objective of these numerical analyses is to validate the methodology described to establish the mechanical and elastic properties of the rPET in the numerical software and to contrast the numerical results obtained with the results obtained after carrying out the experimental tests. The rPET has been defined as an isotropic, elastic, and linear material, with a Poisson's ratio of magnitude 0.38. For Young's modulus, $E_c$, the magnitudes of this parameter obtained from the experimental tests have been used (see Table 5, Table 6, Table 7, and Table 8).

Next, for the meshing process of the geometry, three-dimensional elements of the SOLID 92 type have been defined. This type of finite element is second order and is made up of 10 nodes (see Fig. 18). Of which, 4 nodes are located at the vertices of the tetrahedron and 6 nodes are located at the midpoints of its edges. The precision used in the meshing process is 2.5 mm. Finally, Table 10 and Fig. 18 show the main characteristics of the meshing process defined to perform the numerical simulations.

Table. 10 Mesh statistics defined for the numerical simulations

| Number of elements | Number of nodes | Element quality (average) |
|---|---|---|
| 43,133 | 65,804 | 0.823 |

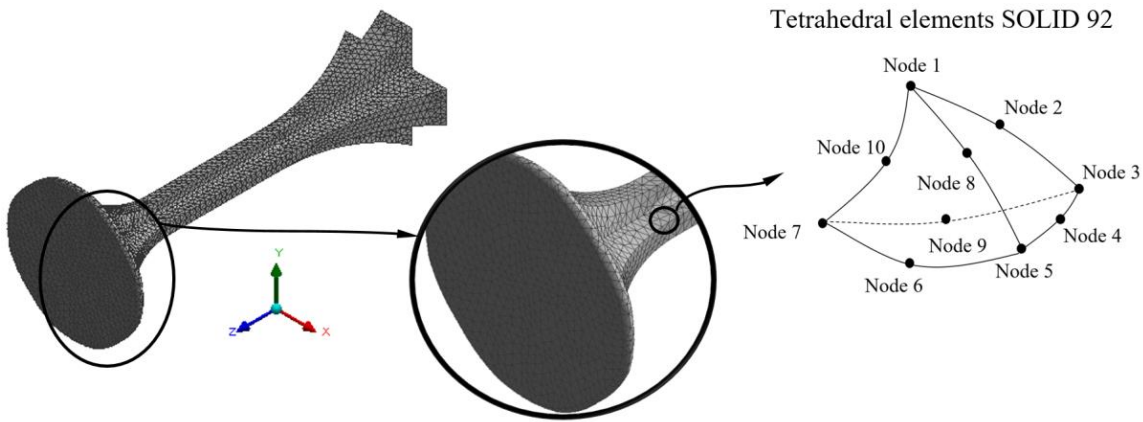

Fig. 18 Mesh definition for the numerical simulation of the design stool

## 6. Results and discussion

According to the structural requirements of the design (see Fig. 2), the serviceability limit state that must be met is defined by a maximum uniaxial compression force of magnitude greater than 1000 N. As can be seen, and based on the obtained experimental results (see Table 9 and Fig. 12), the prototypes of the stool comply with the required requirements, since the maximum uniaxial compression forces that they admit present a magnitude of 6532 N, 5813 N, 5615 N, 5243 N, 6440 N, 5271 N, 6869 N and 5263 N (see Table 9), respectively for each of the different 3D additive manufacturing configurations studied (see Fig. 4). For this reason, it is shown that the structural safety of the stool is not compromised, in any case, by the load scenario and the boundary conditions to which they are subjected. Furthermore, despite carrying out the experimental tests with prototypes subject to a reduced scale, these comply with the established mechanical requirements, and, therefore, the full-scale stool will also comply with them. Furthermore, it is possible to validate the use of the material plastic rPET as well as the manufacturing MEX process for the studied stool according to the presented mechanical and technological requirements. On the other hand, comparing the experimental results and the numerical results is one of the objectives established. As well as, to validate the modeling used to define the elastic and mechanical properties of the rPET in the numerical software for carrying out the mechanical simulations.

Fig. 19, Fig. 20, Fig. 21, and Fig. 22 show the field of nominal displacements obtained from the different numerical simulations carried out, for a load scenario composed of a uniaxial compression force of magnitude 1000 N and a direction aligned with the longitudinal axis of the geometry, and fixed support boundary conditions or embedment applied on the basis. As it is shown in Table 11, Fig. 19, Fig. 20, Fig. 21, and Fig. 22, the relation between numerical results and experimental results presents a percentage of relative error between 0.10% ~5.91% and an absolute error between 0.001 mm~0.024mm. These results validate the numerical model as an evaluation tool for rPET and the printed stool's structural behavior.

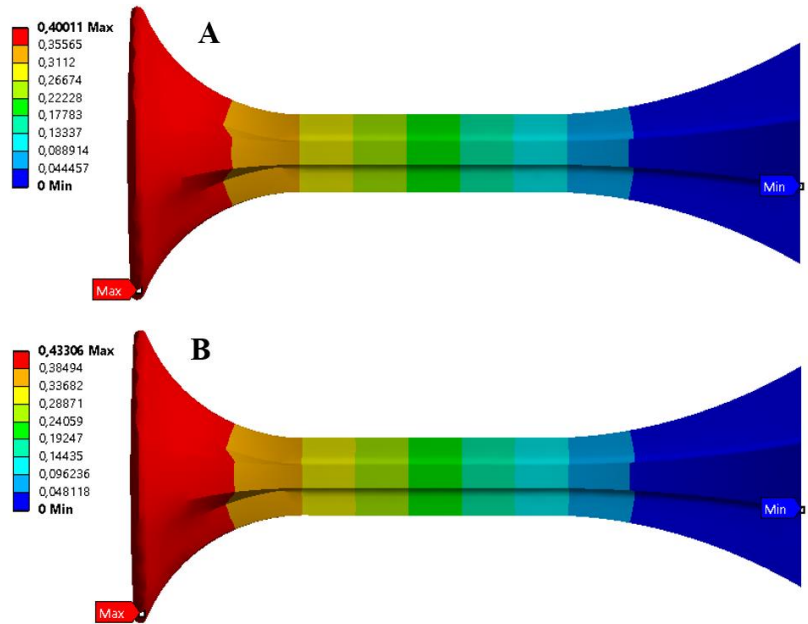

**Fig. 19** Nominal displacement [mm]. a) Concentric Z axis. b) Concentric X/Y axis

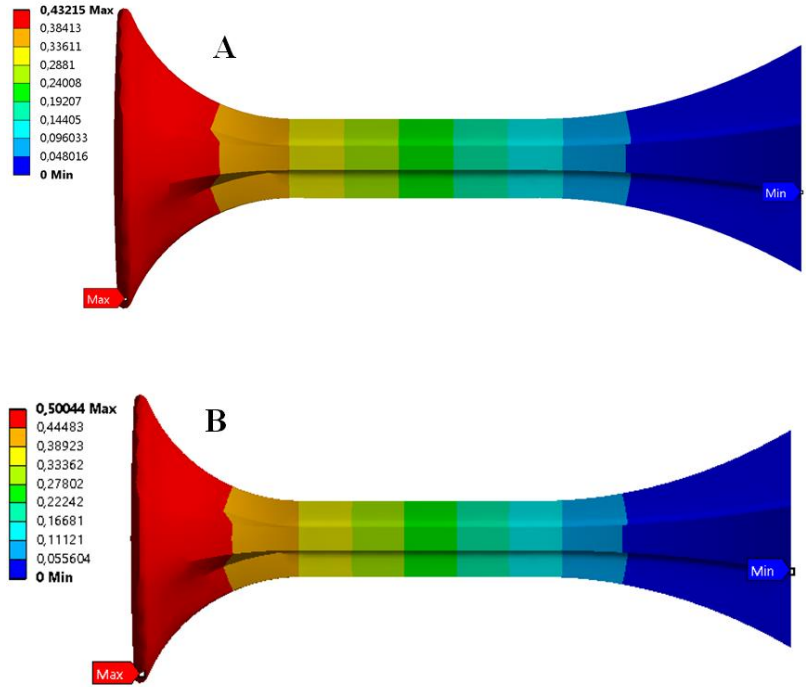

**Fig. 20** Nominal displacement [mm]. a) Zig – Zag Z axis. b) Zig – Zag X/Y axis

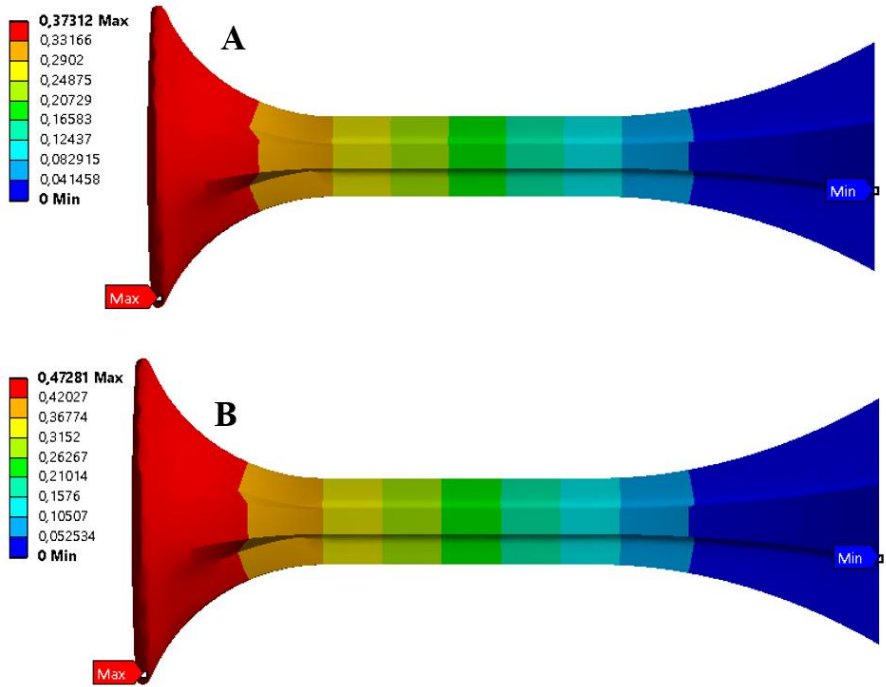

**Fig. 21** Nominal displacement [mm]. a) θ = -45º/45º Z axis. b) θ = – 45º/45º X/Y axis

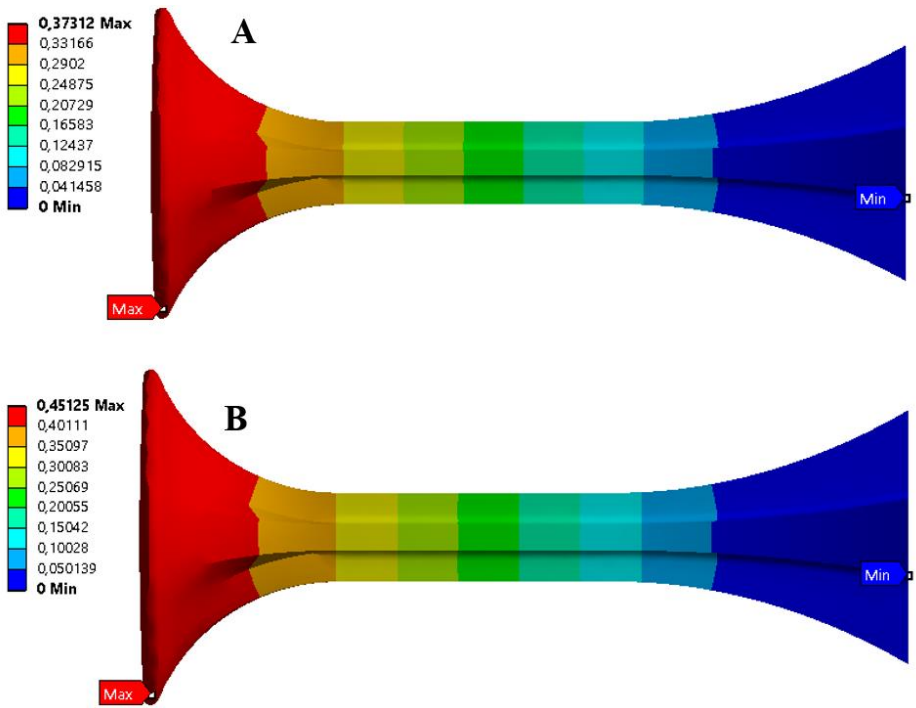

**Fig. 22** Nominal displacement [mm]. a) θ = 0º/90º Z axis. b) θ = 0º/90º X/Y axis

Table. 11 Experimental and numerical results obtained for the design

| Concentric | Experimental displacement [mm] | Numerical displacement [mm] | Absolute error [mm] | Relative error [%] |
|---|---|---|---|---|
| Z | 0.408 | 0.400 | 0.008 | 1.81 |
| X/Y | 0.442 | 0.433 | 0.009 | 2.02 |
| **Zig - Zag** | | | | |
| Z | 0.458 | 0.432 | 0.035 | 5.24 |
| X/Y | 0.502 | 0.500 | 0.002 | 0.31 |
| **θ = – 45º/45º** | | | | |
| Z | 0.399 | 0.375 | 0.024 | 5.91 |
| X/Y | 0.451 | 0.472 | 0.021 | 4.63 |
| **θ = 0º/90º** | | | | |
| Z | 0.399 | 0.375 | 0.024 | 5.91 |
| X/Y | 0.451 | 0.451 | 0.001 | 0.10 |

On the other hand, Fig. 23 shows the magnitude and distribution of the Von – Mises stress field along the surface of the stool. The magnitude of this parameter is directly associated with the geometry and topology analyzed, as well as with the boundary conditions and load scenario to which it is subjected. That is to say, the Von – Mises stress field is analogous to all the 3D printing configurations used since the numerical simulations assume that the plastic material of the stool is isotropic and the analysis mechanical is static and linear. The maximum magnitude of Von – Mises stress obtained is equal to 6.757 MPa, which, as can be verified from the elastic and mechanical properties of the rPET obtained from the experimental tests (see Table 5, Table 6, Table 7, and Table 8), it moves away from the magnitudes of Yield stress, $\sigma_{yc}$, and the ultimate stress, $\sigma_{uc}$. So, for the defined load scenario and boundary conditions, the structural behavior of the stool is completely elastic and linear.

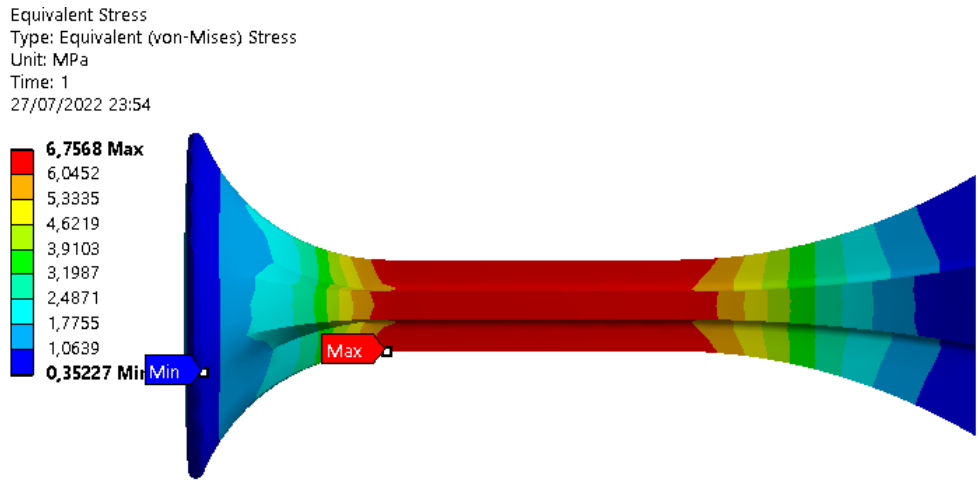

Fig. 23 Von – Mises stress

Analyzing the elastic properties obtained experimentally according to the international standard ASTM D695-15, it is concluded that the additively manufactured rPET using a MEX process can be qualified isotropically in its modeling for use in numerical simulations and under compressive loads. The main advantage lies in the fact that to perform a numerical simulation it is not necessary to change the material properties in the validation software. Additionally, it is not necessary to select the orientation of the material in the simulation software, always taking into account that the design works within the elastic zone. Outside this zone, the material is anisotropic, so manufacturing properties are required.

## 7. Conclusions

This paper presents the numerical and experimental modeling of the mechanical behavior of the recycled polymer rPET subjected to gravitational compressive loads for additive MEX manufacturing applications in sustainable products. To reach the objectives established, the research work has focused on the application of two types of experimental tests. Firstly, the structural and mechanical behavior of the different configurations of rPET test specimens manufactured using MEX technology has been experimentally analyzed. From the experimental results, it is verified that under pure compression stresses the printing configurations analyzed in the X/Y and Z axes present similarities in terms of the values of the elastic modulus, highlighting the Z direction as the one that shows the best elastic behavior. In particular, the average value of the Young's modulus obtained for the cylindrical specimens manufactured in the printing direction Z is equal to 1,952 MPa, which supposes an improvement of 20% in the elastic behavior of the rPET plastic material, with respect to the results obtained for the print direction X/Y. Furthermore, and based on the experimental results obtained for the 3D patterns considered, it is determined that the pattern θ = 0º/90º is the configuration that presents a greater magnitude of Young's modulus parameter under compression stresses and therefore the best elastic and structural behavior compared to the rest of the printing patterns. Since the values obtained for the Young's modulus are equal to 2,089 MPa and 1,688 MPa, for the printing direction X/Y and Z respectively, exceeding the values obtained for the pattern θ = – 45º/45º, 2,089 MPa and 1,549 MPa respectively, the values obtained for the Zig – Zag pattern, 1,849 MPa and 1,543 MPa respectively, and the values obtained for the Concentric pattern, 1,781 MPa and 1,744 MPa respectively. All of this implies an improvement in the elastic behavior of the rPET material for the θ = 0º/90º pattern of 4%, 11% and 7% compared to the θ = – 45º/45º, Zig–Zag and concentric patterns, respectively. In this line, a higher yield strength value is observed in the elastic zone for configurations build along the X/Y directions, except the pattern θ = – 45º/45º. Specifically, the average value of the yield strength obtained for the cylindrical specimens manufactured in the printing direction X/Y is equal to 37.17 MPa, which supposes an improvement of 10% in the elastic limit of the rPET plastic material, with respect to the results obtained for the print direction Z. This is because, under compressive stresses, the material exhibits structural stiffness until reaching the elastic limit of the extruded filaments, the point from which the additively manufactured material collapses. On the contrary, the configurations manufactured along the Z axis present a better elastic behavior including a lower value of yield strength and a lower mechanical stiffness to the material manufactured in this direction. The analysis in the plastic zone of the stress-strain curves presents a flat area indicative of an increase in strain versus a constant load. This is because the material, after presenting an elastic behavior, undergoes a plastic deformation process, in which the gaps between the peaks and valleys of the weld beads are filled reaching a hardening of the manufactured material, collapsing finally in the same way to a ductile material. Once the mechanical and elastic characterization of rPET has been completed using test specimens, eight experimental tests of the designed real sustainable part have been carried out following configurations analogous to those defined for the cylindrical test specimens for the 3D additive manufacturing process. The results of the experimental tests for the industrial part are analogous to those obtained in the tests for the test specimens.

Next, and as a fundamental objective of the paper, the additively manufactured rPET has been numerically characterized for its use in numerical validations of eco-products. The objective has been to evaluate the influence of the anisotropy in the numerical analysis of rPET additively manufactured under gravitational compression loads. The paper presented confirms that it is possible to carry out the simulations considering the rPET material as isotropic in the elastic area, given that after characterizing the compression material experimentally in accordance with the standards and validating an example of complex geometry, it shows that the maximum deviation obtained after the analysis in all the cases studied does not exceed 5%. This finding is tremendously important given that high-level industrial design companies are beginning to use recycled materials and additive manufacturing processes for their respect for the environment. Although there are several papers [Zander et al. 2018; Bex et al. 2021; Vidakis et al. 2021] related to the mechanical characterization of the rPET material for MEX process there is currently no paper demonstrating this results for rPET material. However, for the working range outside the elastic area, the material behaves anisotropically, so it will be necessary to consider the particular properties for each manufacturing requirement in the numerical model of the material.

Eight numerical simulations have been carried out to analyze the structural behavior of the geometry for each of the different 3D additive manufacturing configurations. Comparing the experimental results and the numerical results, an absolute error between 0.001 mm ~ 0.024 mm has been obtained. These results validate the numerical model as an evaluation tool for rPET and the printed stool 's structural behavior.

The results presented in the paper validate the use of the additively manufactured rPET material for industrial use and in sustainable consumer products subjected to compressive stresses. Major design companies such as (Kartell 2022) are now beginning to use recycled plastic materials in their high-end designs. The results of the paper demonstrate the feasibility of using the rPET material and the MEX process for sustainable industrial products using recycled materials and a sustainable manufacturing process such as the additive. Additionally, the use of rPET material has been characterized for its use in Finite Element Analysis (FEA) applications. This information has a great impact at the research and industrial level since this information is not available to date, making it impossible to numerically validate a product manufactured with an rPET material due to the inherent anisotropy and uncertain qualities of 3D printed parts manufactured with MEX processes.

**Appendix**

Figure 24 represents the workflow for filament fabrication and 3D printing (MEX process) .

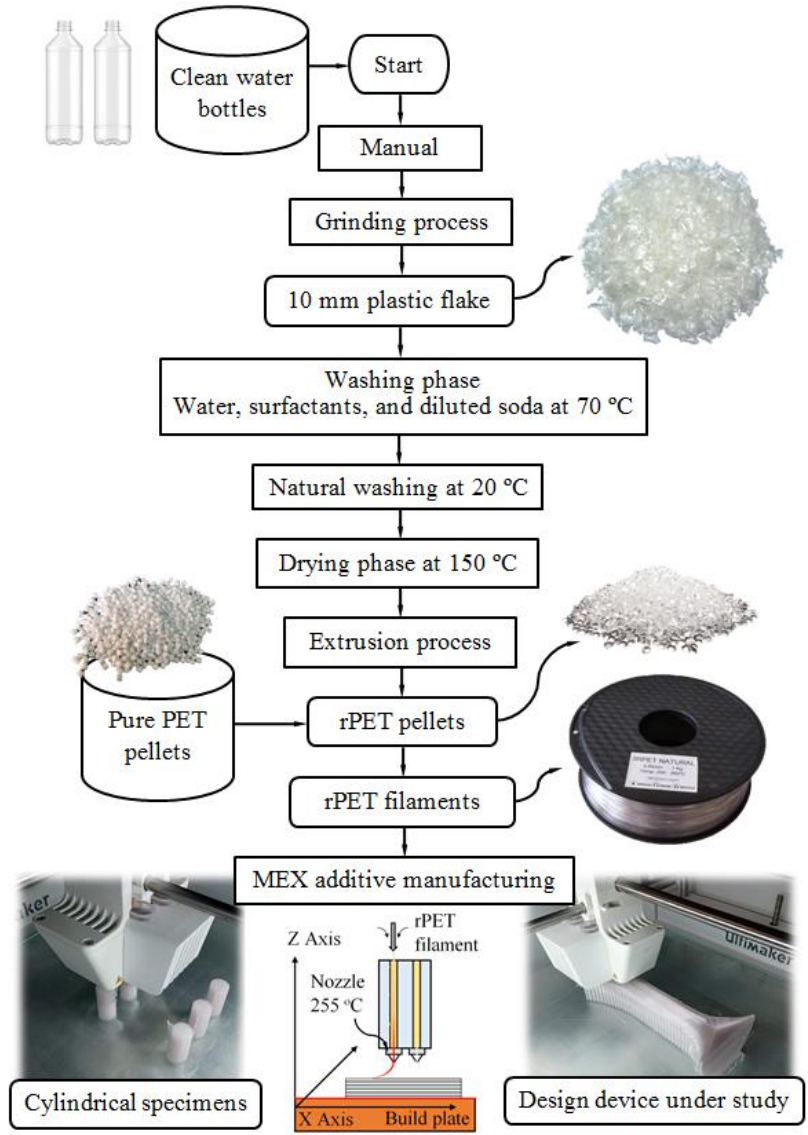

Fig. 24 Workflow for filament fabrication and MEX additive manufacturing

**Funding**

This research work was supported by the University of Jaen through the Plan de Apoyo a la Investigación 2021–2022-ACCION1a POAI 2021–2022: TIC-159

**Conflict of Interest**

The authors declare that they have no conflict of interest. Figures and Tables by authors